\documentclass[10pt,twocolumn]{article}
\usepackage[top=2cm, bottom=2cm, left=1.8cm, right=1.8cm]{geometry}
\usepackage{times}  %
\usepackage[hyphens]{url}
\usepackage{graphicx}  %
\usepackage[keeplastbox]{flushend}
\usepackage{xcolor}
\usepackage{booktabs}
\usepackage{graphicx}
\usepackage{paralist}
\usepackage[small,bf]{caption}
\usepackage[tight]{subfigure}
\usepackage[hang,flushmargin]{footmisc}
\renewcommand{\footnotesize}{\fontsize{8}{9}\selectfont}
\usepackage[compact]{titlesec}
\titlespacing*{\section}{0pt}{*4}{4pt}
\titlespacing{\subsection}{0pt}{*3}{3pt}
\usepackage{xspace}

\makeatletter
\def\url@leostyle{%
  \@ifundefined{selectfont}{\def\UrlFont{}}%
  {\def\UrlFont{}}%
}
\makeatother
\usepackage[hyphenbreaks]{breakurl}

\usepackage{ifthen}
\newif\ifshort
\shorttrue   %
\ifshort
  \newcommand{\isShort}{true}
\else
  \newcommand{\isShort}{false}
\fi
\newcommand{\shortVer}[1]{\ifthenelse{\equal{\isShort}{true}}{{#1}}{}}
\newcommand{\longVer}[1]{\ifthenelse{\equal{\isShort}{false}}{{#1}}{}}

\newcommand{\descr}[1]{\smallskip\noindent\textbf{#1}}

\newcommand{\dspol}{{/pol/}\xspace}

\usepackage{url}

\begin{document}
\title{\bf Characterizing the Use of Images in State-Sponsored Information Warfare Operations by Russian Trolls on Twitter\thanks{To appear at the 14th International AAAI Conference on Web and Social Media (ICWSM 2020) -- please cite accordingly. Work done while first author was with Cyprus University of Technology.}}

\author{Savvas Zannettou$^{1}$, Tristan Caulfield$^2$, Barry Bradlyn$^3$,\\Emiliano De Cristofaro$^2$, Gianluca Stringhini$^4$, and Jeremy Blackburn$^5$\\[0.5ex]
\normalsize $^1$Max Planck Institute for Informatics, $^2$University College London, $^3$University of Illinois at Urbana-Champaign,\\
\normalsize$^4$Boston University, $^5$Binghamton University}
\date{}

\maketitle

\begin{abstract}

State-sponsored organizations are increasingly linked to efforts aimed to exploit social media for information warfare and manipulating public opinion.
Typically, their activities rely on a number of social network accounts they control, aka trolls, that post and interact with other users disguised as ``regular'' users.
These accounts often use images and memes, along with textual content, in order to increase the engagement and the credibility of their posts.

In this paper, we present the first study of images shared by state-sponsored accounts by analyzing a ground truth dataset of 1.8M images posted to Twitter by accounts controlled by the Russian Internet Research Agency.
First, we analyze the content of the images as well as their posting activity.
Then, using Hawkes Processes, we quantify their influence on popular Web communities like Twitter, Reddit, 4chan's Politically Incorrect board (\dspol), and Gab, with respect to the dissemination of images.
We find that the extensive image posting activity of Russian trolls coincides with real-world events (e.g., the Unite the Right rally in Charlottesville), and shed light on their targets as well as the content disseminated via images.
Finally, we show that the trolls were more effective in disseminating politics-related imagery than other images.

\end{abstract}

\section{Introduction}\label{sec:intro}

Social network users are constantly bombarded with digital content. %
While the sheer amount of information users have access to was unthinkable just a couple of decades ago, %
the way in which people process that information has also evolved drastically.
Social networks have become a battlefield for \emph{information warfare}, with different entities attempting to disseminate content to achieve strategic goals, push agendas, or fight ideological battles~\cite{rowett2018StrategicNeedUnderstand,denning1999information}.

As part of this tactic, governments often employ ``armies'' of actors, operating from believable accounts and posting content that aims to manipulate opinion or sow public discord by actively participating in online discussions.
Previous work has studied the involvement of state-sponsored accounts in divisive events, e.g., the Black Lives Matter movement~\cite{stewart2018examining} or the 2016 US elections~\cite{badawy2018analyzing,boyd2018characterizing}, %
highlighting how these entities can be impactful both on the information ecosystem and in the real world. %

In today's information-saturated society, the effective use of images when sharing online content can have a strong influence in whether content will catch people's attention and go viral~\cite{berger2012makes,jenders2013analyzing,khosla2014makes}.
Users often feel overwhelmed with how much content they are exposed to~\cite{koroleva2010stop}, and pay attention to each piece of information for short amounts of time, with repercussion to their attention span~\cite{wrzus2013social}.
In fact, previous research showed that 60\% of social network users re-share articles on social media without reading them, basing their decision on limited cues such as the title of the article or the thumbnail image associated with it~\cite{gabielkov2016social}.

\begin{figure}[t]
\centering
\includegraphics[width=\columnwidth]{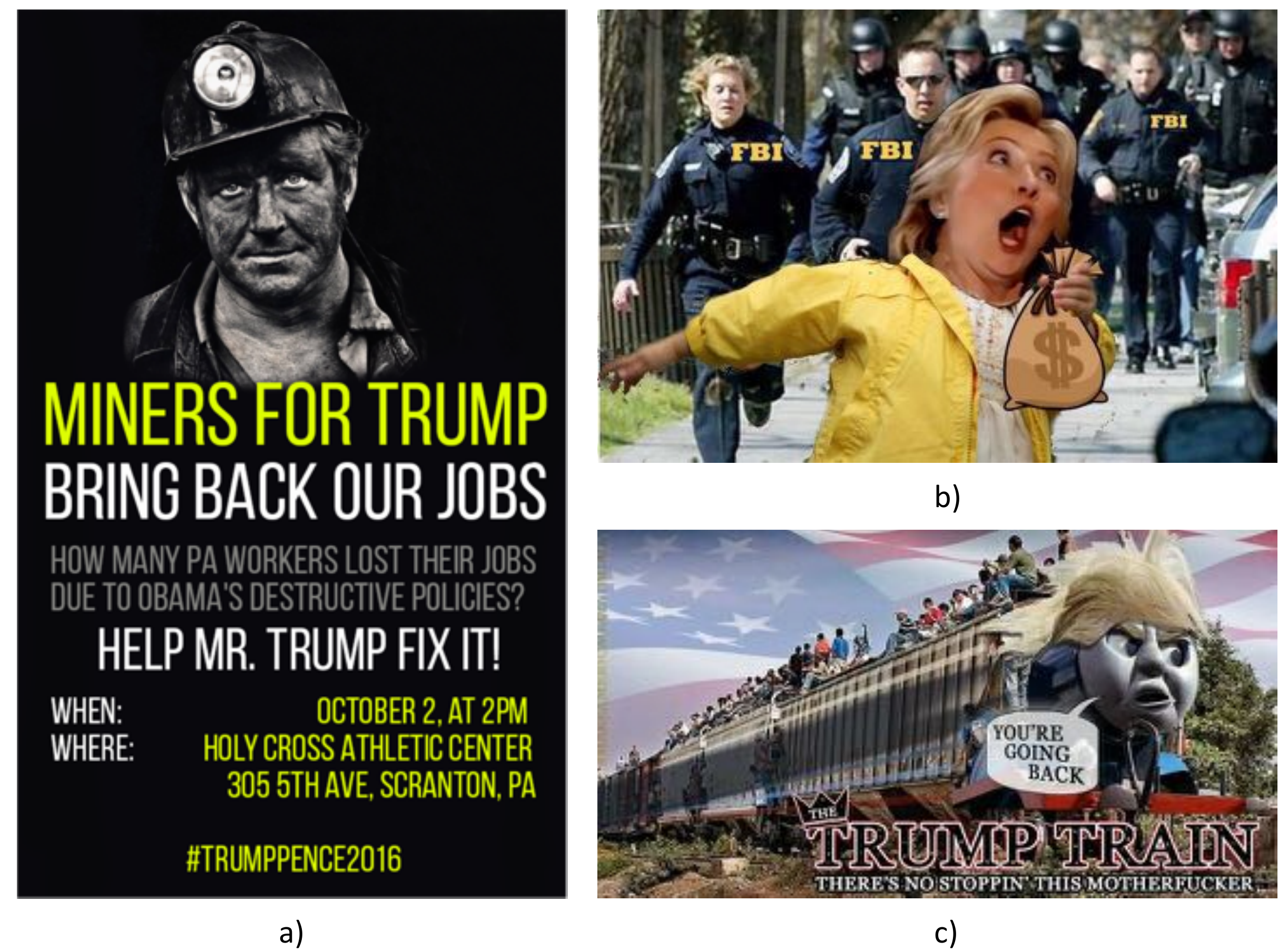}
\vspace{-0.5cm}
   \caption{Examples of politically-charged images posted by Russian trolls.}
\label{fig:motivation_images}
\vspace{-0.25cm}
\end{figure}

Therefore, as part of the efforts aimed to actively push agendas, state-sponsored accounts do not only use textual content, but also take advantage of the expressive power of images and pictures,
e.g., using politically and ideologically charged memes~\cite{rowett2018StrategicNeedUnderstand}. %
In Figure~\ref{fig:motivation_images}, we report some (self-explanatory) examples of images pushed by state-sponsored accounts on Twitter, showcasing their unequivocally political nature and how they can be used to push agendas.
Nonetheless, the role of images in information diffusion on the Web has attracted limited attention from the research community, which has thus far mainly focused on textual content~\cite{badawy2018analyzing}. %

In this paper, we begin filling this gap by studying the use of images by state-sponsored accounts, aka Russian trolls~\cite{twitter_russian_iranians_dataset}.
In particular, we focus on the following research questions:
\begin{compactenum}
\item What content is disseminated via images by state-sponsored accounts?
\item Can we identify the target audience of Russian state-sponsored accounts by studying the images they share?
\item How influential are these accounts in making images go viral on the Web? How does this influence results compare to previous characterizations that look into the spread of news by these accounts? %
\end{compactenum}

Aiming to address these questions, we use an image-processing pipeline, expanding that presented by~\cite{zannettou2018origins}, to study images shared by state-sponsored trolls on Twitter.
More precisely, we implement a custom annotation module that uses Google's Cloud Vision API to annotate images in the absence of high-quality ground truth data, or for images that are not bounded to a specific domain (e.g., memes).
We then run the new pipeline on a dataset of 1.8M images from the 9M tweets released by Twitter in October 2018 as part of their effort to curb state-sponsored propaganda~\cite{twitter_russian_iranians_dataset}.
These tweets were posted by 3.6K accounts
identified as being controlled by the Russian Internet Research Agency (IRA).
Finally, we quantify the influence that state-sponsored trolls had on other mainstream and alternative Web communities: namely, Twitter, Reddit, Gab, and 4chan's Politically Incorrect board (\dspol).
To do this, we use Hawkes Processes~\cite{linderman2014,lindermanArxiv}, %
which allow us to model the spread of the images across multiple Web communities and assess the root cause of the image appearances.

\descr{Main Findings.} Along with a first-of-its-kind characterization of how images are used by state-sponsored actors, our work yields a number of interesting findings:
\begin{compactenum}
\item The sharing of images by the trolls coincides with real-world events.
For instance, we find a peak in activity that is clearly in close temporal proximity with the Unite the Right rally in Charlottesville~\cite{charlotesville}, likely suggesting their use to sow discord during dividing events. %
\item Our analysis provides evidence of their general themes and targets. %
For instance, we find that Russian trolls were mainly posting about Russia, Ukraine, and the USA.
\item By studying the co-occurrence of these images across the Web, we show that the same images appeared in many popular social networks, as well as mainstream and alternative news outlets. Moreover, %
we highlight interesting differences in popular websites for each of the detected entities: for instance, troll-produced images related to US matters were mostly co-appearing on mainstream English-posting news sites.

\item Our influence estimation results highlight that the Russian state-sponsored trolls, despite their relatively small size, are particularly influential and efficient in pushing images related to politics to other Web communities.
In particular, we find that Russian state-sponsored trolls were more influential in spreading political imagery when compared to other images.
Finally, by comparing these results to previous analysis focused on news~\cite{zannettou2018let}, we find that trolls were slightly more influential in spreading news via URLs than images.
\end{compactenum}

\section{Related Work}

\descr{Trolls and politics.} Previous work has focused on understanding the behavior, role, and impact of state-sponsored accounts on the US political scene.
Boyd et al.~\cite{boyd2018characterizing} perform linguistic analysis on posts by Russian state-sponsored accounts over the course of the 2016 US election; they find that right- and left-leaning communities are targeted differently to maximize hostility across the political spectrum in the USA.
Stewart et al.~\cite{stewart2018examining} investigate the behavior of state-sponsored accounts around the BlackLivesMatter movement, finding that they infiltrated both right- and left-leaning political communities to participate in both sides of the discussions.
Jensen~\cite{jensen2018russian} finds that, during the 2016 US election, Russian trolls were mainly %
interested in defining the identity of political individuals rather than particular information claims.

\descr{Trolls in social networks.} %
Dutt et al.~\cite{dutt2018senator} analyze the advertisements purchased by Russian accounts on Facebook.
By performing clustering and semantic analysis, they identify their targeted campaigns over time, concluding that their main goal is to sway division on the community, and also that the most effective campaigns share similar characteristics.
Zannettou et al.~\cite{zannettou2018disinformation} compare a set of Russian troll accounts against a random set of Twitter users, showing that Russian troll accounts exhibit different behaviors in the use of the Twitter platform when compared to random users.
In follow up work, Zannettou et al.~\cite{zannettou2018let} analyze the activities of Russian and Iranian trolls on Twitter and Reddit, finding substantial differences between them (e.g., Russian trolls were pro-Trump, Iranian ones anti-Trump), that their behavior and targets vary greatly over time, and that Russian trolls discuss different topics across Web communities (e.g., they discuss about cryptocurrencies on Reddit but not on Twitter).
Also, Spangher et al.~\cite{spangher2018analysis} examine the exploitation of various Web platforms (e.g., social networks and search engines), showing that state-sponsored accounts use them to advance their propaganda by promoting content and their own controlled domains.
Finally, Broniatowski et al.~\cite{broniatowski2018weaponized} focus on the vaccine debate and study  Twitter discussions by Russian trolls, bots, and regular users.
They find that the trolls amplified both sides of the debate, while at the same time their messages were more political and divisive in nature when compared to messages from bots and regular users.

\begin{figure}[t]
\centering
\includegraphics[width=\columnwidth]{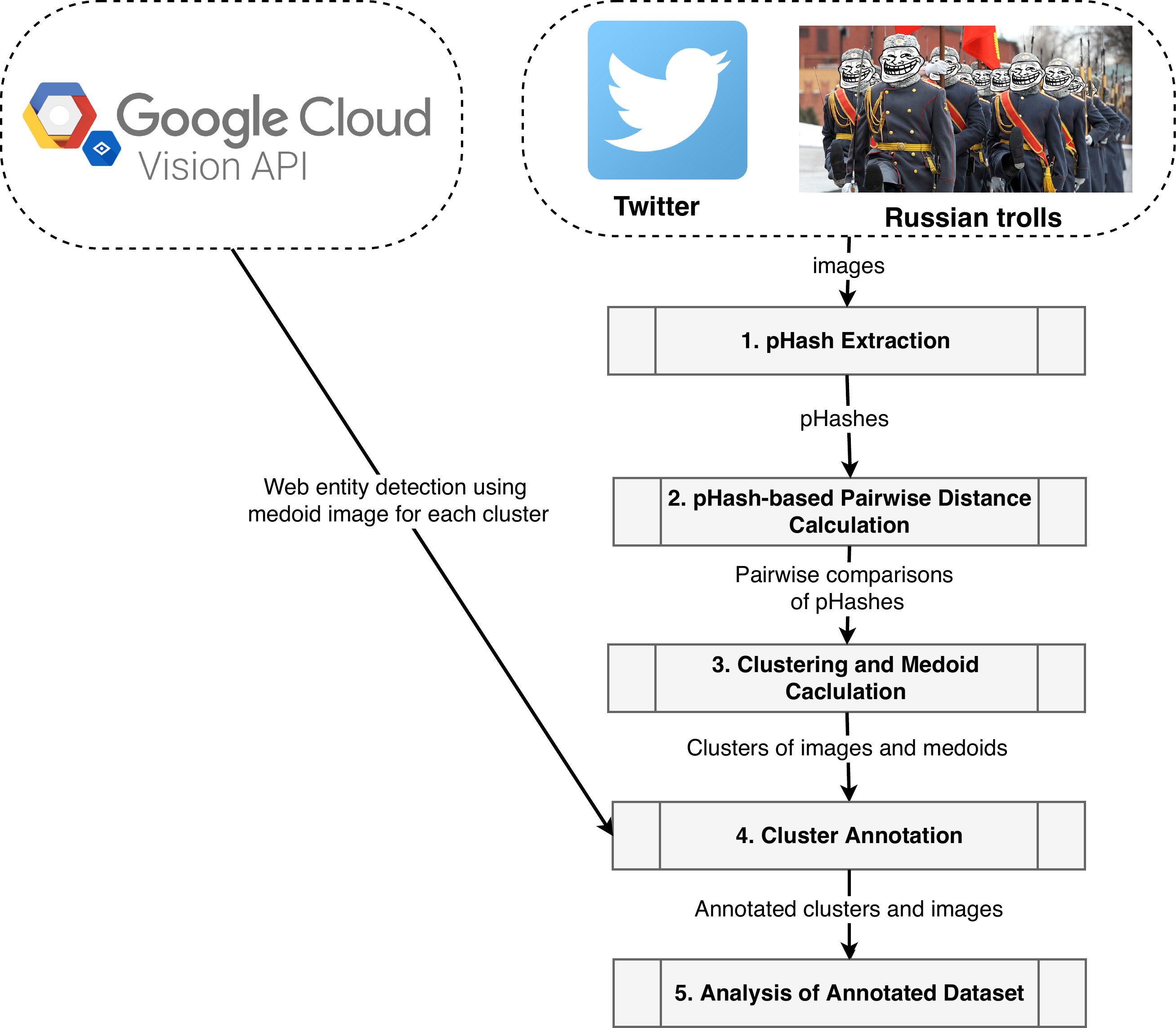}
   \caption{Overview of our image processing pipeline.} %
\label{fig:pipeline}
\end{figure}

\descr{Detection \& Classification.} Badawy et al.~\cite{badawy2018falls} use machine learning to detect Twitter users that are likely to share content that originates from Russian state-sponsored accounts, while Im et al.~\cite{im2019still} detect Russian trolls using machine learning techniques, finding that these accounts are still very active on the Web.
Also, Kim et al.~\cite{kim2019tracking} classify Russian state-sponsored trolls into various roles: left- or right-leaning or accounts that pose as news outlets.
By applying their technique on 3M tweets posted by Russian trolls on Twitter, they find that despite the fact that trolls had multiple roles, they worked together, while for trolls that pose as news outlets, they find that they had multiple agendas.
For instance, some were posting about violent news to create an atmosphere of fear, while others focused on posting highly biased political news.

\descr{Remarks.} Overall, unlike previous work, we focus on content shared via {\em images} by state-sponsored accounts.
Indeed, to the best of our knowledge, ours is the first study performing a large-scale image analysis on a ground truth dataset of images shared by Russian trolls on Twitter.
Previous research~\cite{gabielkov2016social} has showed that social network users usually decide what to share and consume content based on visual cues; thus,  as state-sponsored accounts tend to post disinformation~\cite{mejias2017disinformation}, studying the images they share provides an important tool to understand and counter disinformation.

\section{Methodology}\label{sec:methodology}
We now present our dataset and our methodology for analyzing images posted by state-sponsored trolls on Twitter.

\shortVer{
\descr{Dataset.} We use a ground truth dataset of tweets posted by Russian trolls released by Twitter in October 2018~\cite{twitter_russian_iranians_dataset}.
The dataset includes over 9M tweets posted by 3.6K Russian state-sponsored accounts, and their associated metadata and media (1.8M images).
Note that the methodology employed by Twitter for detecting/labeling these state-sponsored accounts is not publicly available. %
That said, to the best of our knowledge, this is the most up-to-date and the largest ground truth dataset of state-sponsored accounts and their activities on Twitter.
}

\descr{Ethics.} We only work with publicly available data, which was anonymized by Twitter, and follow standard ethical guidelines~\cite{rivers2014ethical}---e.g., we do not try to de-anonymize users based on their tweets.

\descr{Image analysis pipeline.}
To analyze the images posted by these state-sponsored accounts, we build on the image processing pipeline presented by~\cite{zannettou2018origins}.
This relies on Perceptual Hashing, or pHash~\cite{monga2006perceptual}, and clustering techniques~\cite{ester1996density} to group similar images according to their visual peculiarities, yielding clusters of visually similar images.
Then, clusters are annotated based on the similarity between a ground truth dataset and each cluster's medoid (i.e., the representative image in the cluster).
For this process, \cite{zannettou2018origins} use crowdsourced meme metadata obtained from Know Your Meme.
Our effort, however, has a broader scope as the images shared by state-sponsored accounts are not limited to memes.
Consequently, we use a different annotation approach, relying on Google's Cloud Vision API\footnote{\url{https://cloud.google.com/vision/}}, a state-of-the-art solution in Computer Vision tasks to gather useful insights from open-domain images (i.e., not bounded to a specific domain like Internet memes).

Figure~\ref{fig:pipeline} shows the resulting extended pipeline.
We perform the ``Web Detection'' task using Cloud Vision API, which provides us with two very useful pieces of information for each image:
1)~a set of \emph{entities}, and their associated confidence scores, that best describe the image (e.g., an image showing Donald Trump yields an entity called ``Donald Trump'');
and 2) a set of \emph{URLs} on the Web that the same image appeared.
To extract this information, the API leverages Google's image search functionality to find URLs to identical and similar images. Furthermore, by extracting data from the text of these URLs, the API provides a set of entities that are related to the image.
These two pieces of information are crucial for our analysis as they allow us to understand the context of the images and their appearance across the Web.

\descr{Running the pipeline.} First, we extract a pHash for each image using the ImageHash library.\footnote{\url{https://github.com/JohannesBuchner/imagehash}}
This reveals that there is a substantial percentage of images that are either visually identical or extremely similar as they have the same pHashes
\shortVer{(43\% of the images)}.
Next, we cluster the images by calculating all the pairwise comparisons of all the pHashes.
This results in 78,624 clusters containing 753,634 images\shortVer{.}
Then, for each cluster, we extract the medoid, which is the image that has the minimum average Hamming distance between all the images in the cluster.
Then, using each medoid, we perform ``Web Detection'' using the Cloud Vision API, which provides us with a set of entities and URLs, which we assign for each image in the cluster.
This is doable since the average number of unique images per cluster is 1.8 with a median of 1 unique image per cluster (see Figure~\ref{fig:cdf_images_per_cluster}).

\descr{Pipeline Evaluation.} To evaluate the performance of our pipeline, we manually annotate a random sample of 500 clusters.
Specifically, the first author of this paper manually checked the 500 random clusters and the corresponding Cloud Vision entity with the highest confidence score to assess whether the entity is ``appropriate'' with respect to the images within the cluster.
We find that the Cloud Vision API-based annotation procedure provides an appropriate entity in 83.7\% of the clusters in the random sample. Thus, we argue this is a reasonable performance for the purposes of our study.

\shortVer{
\begin{figure}[t!]
\center
\subfigure[]{
\includegraphics[width=0.6\columnwidth]{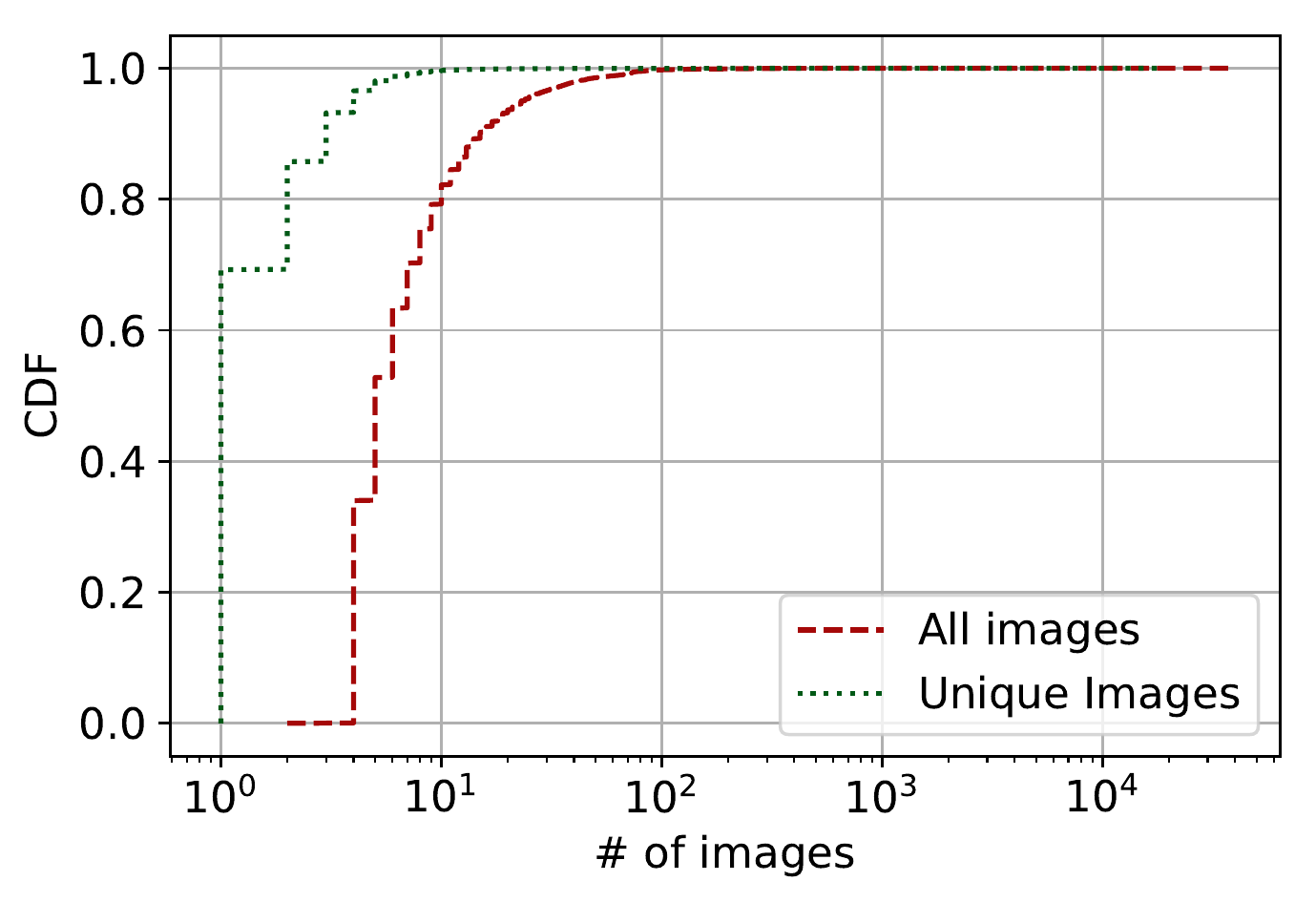}\label{fig:cdf_images_per_cluster}}
\subfigure[]{
\includegraphics[width=0.6\columnwidth]{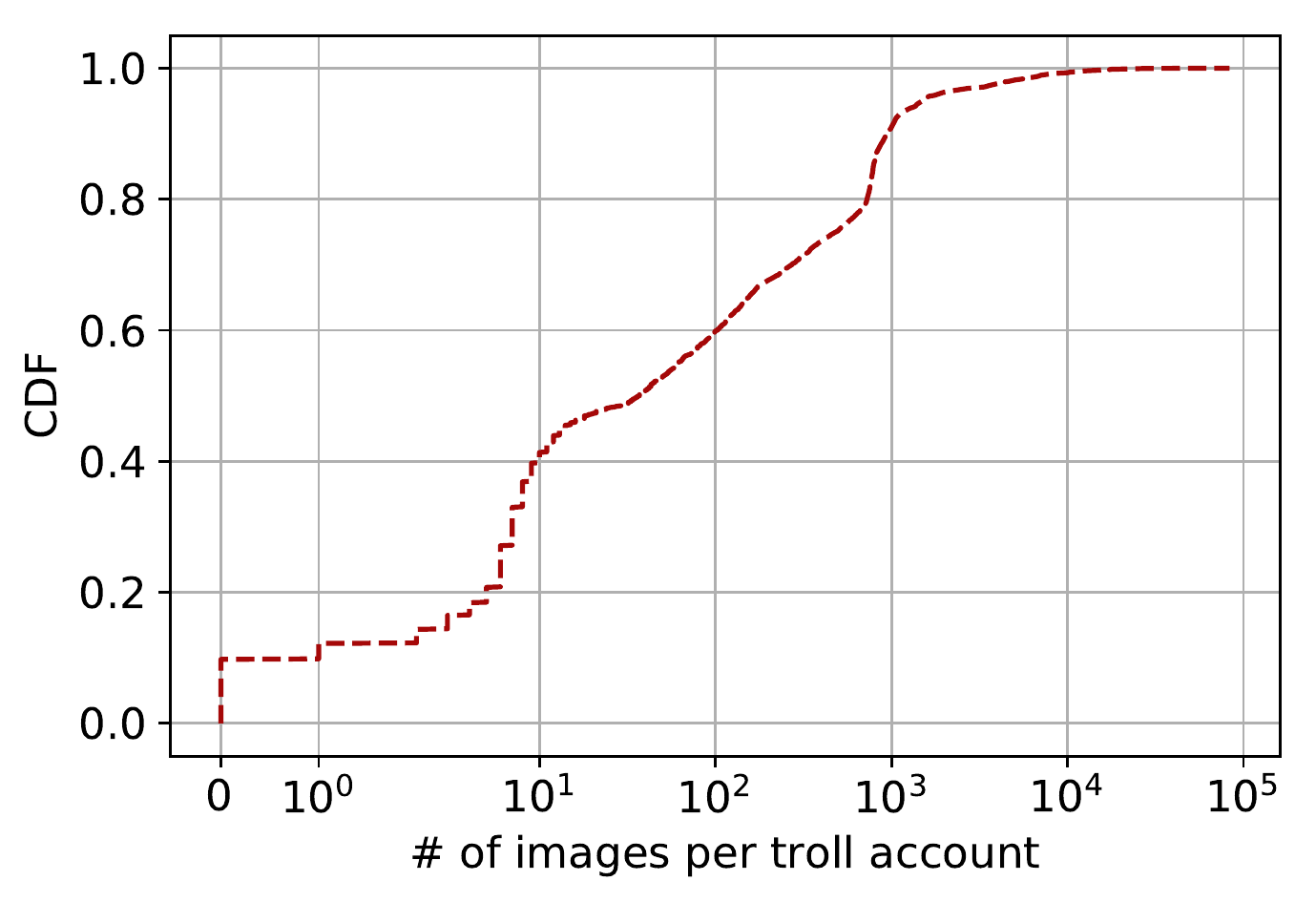}\label{fig:cdf_images_per_user}}
  \caption{CDF of a) number of images per cluster (image uniqueness is based on their pHash); and b) number of images per troll account.}
\label{fig:cdf_images_stats}
\end{figure}
}

\shortVer{
\begin{figure}[t]
\centering
\includegraphics[width=0.6\columnwidth]{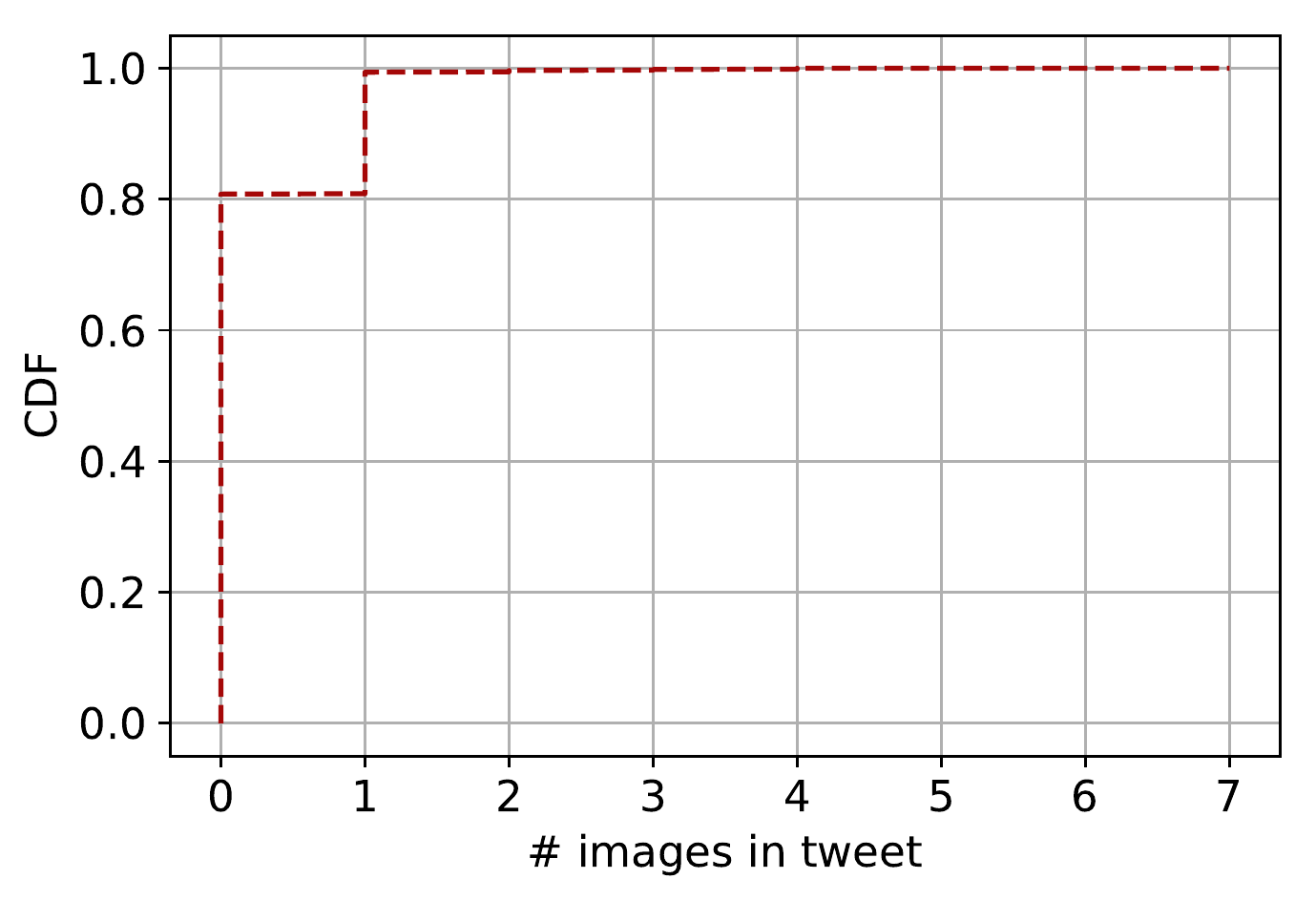}
  \caption{CDF of the number of images per tweet in our dataset.}%
\label{fig:cdf_images_in_tweets}
\end{figure}
}

\section{Image Analysis} \label{sec:results}
We now present the results of our analysis.
First, we perform a general characterization of the images posted by state-sponsored accounts on Twitter and then an analysis of the content of the images.
Also, we study the occurrence of the images across the Web.

\subsection{General Characterization}
We begin by looking at the prevalence of images in tweets by state-sponsored trolls.
In Figure~\ref{fig:cdf_images_per_user}, we plot the CDF of the number of images posted per confirmed state-sponsored account that had at least one tweet (4.5\% of the identified trolls never tweeted). %
We find that only a small percentage of these accounts do not share images \shortVer{(9.7\% of the Russian troll accounts).}
Also, some accounts shared an extremely large number of images, \shortVer{8\% of the Russian trolls posted over 1K images}.
\shortVer{Furthermore, we find an average of 502.2 images per account with a median number of images of 37.}

Then, in Figure~\ref{fig:cdf_images_in_tweets}, we report the CDF of the number of images per tweet; we find that 19\% of tweets posted by Russian trolls include at least one image.
One explanation for this relatively large fraction is that Twitter automatically generates a preview/thumbnail image when you post a URL.
Indeed, by inspecting the URLs in the tweets, we find that out of the 19\% of the tweets that contained images, 11.8\% of them contained automatically generated one, while the rest (7.2\%) include images that are explicitly posted (i.e., not generated based on a posted URL).
That said, we include \emph{all} images in our dataset and analysis, as generated images too provide insight into the content posted by the state-sponsored accounts, especially considering their proclivity to post ``fake news''~\cite{mejias2017disinformation} and the role images might play in catching people's attention.

\begin{figure}[t!]
\centering
\subfigure[]{
\includegraphics[width=\columnwidth]{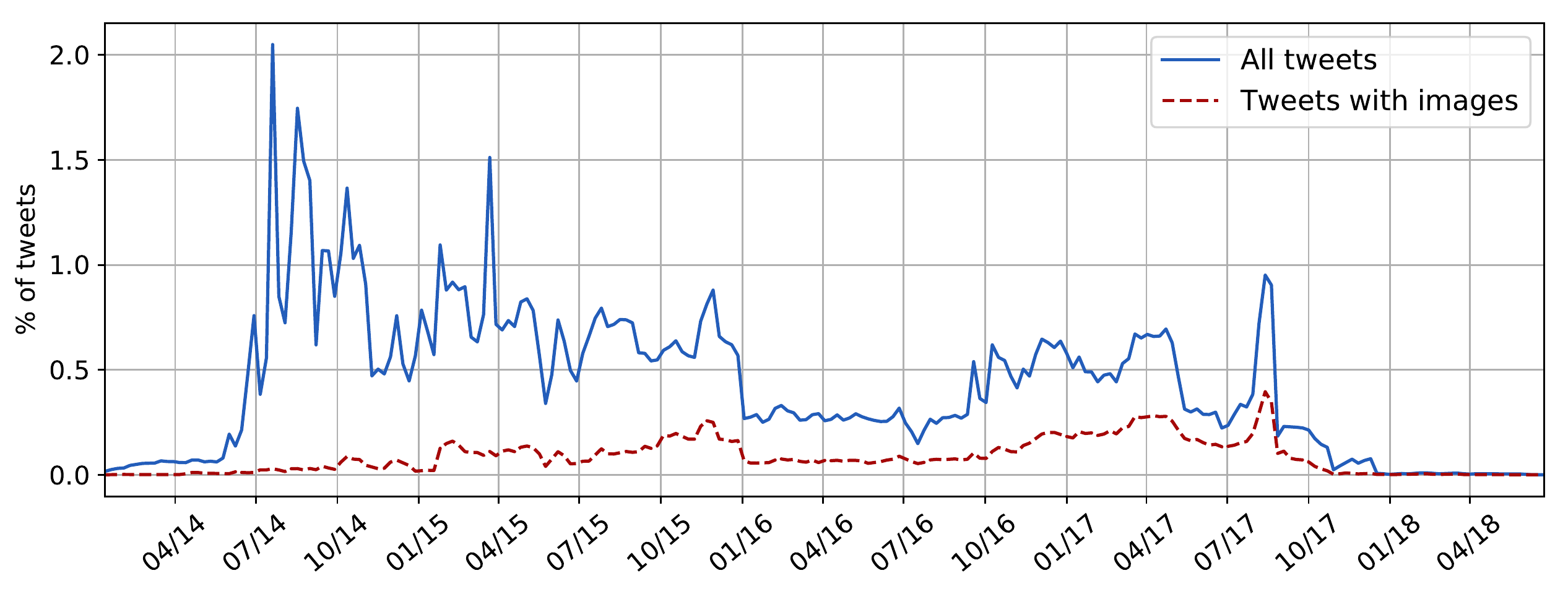}\label{subfig:tweets_temporal}}
\subfigure[]{
\includegraphics[width=\columnwidth]{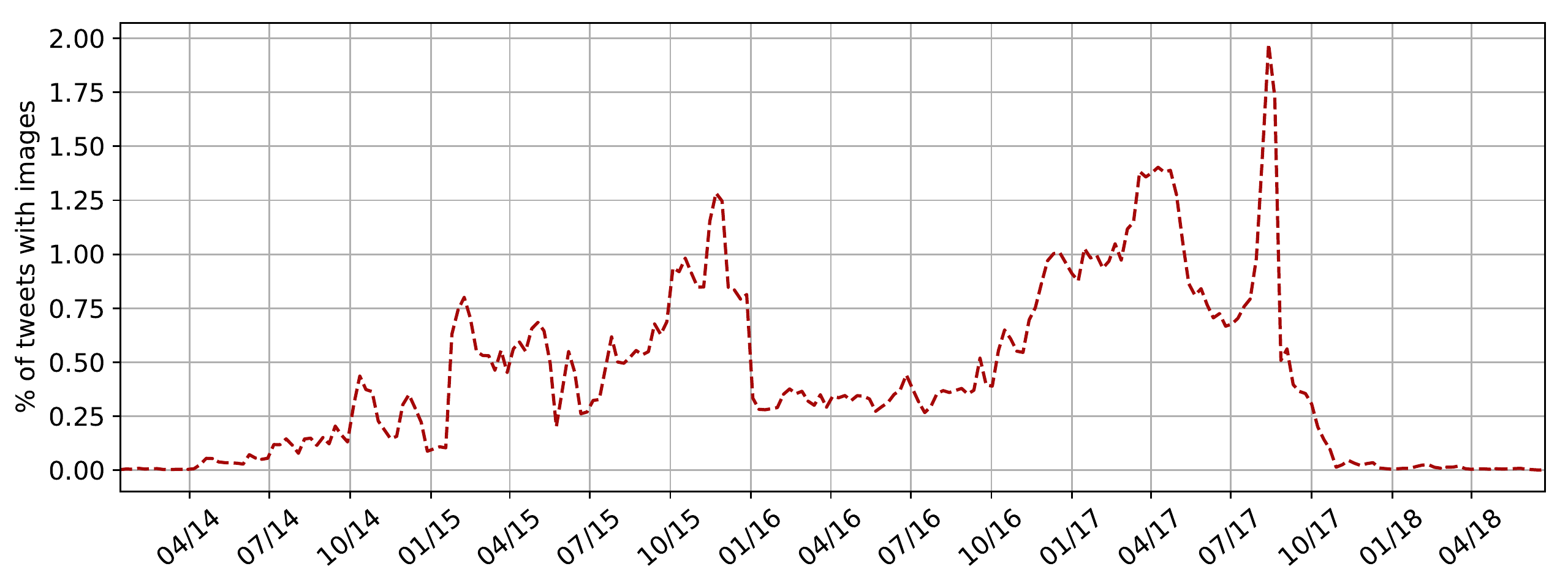}\label{subfig:tweets_images_temporal}}
\caption{Temporal overview of: a) all tweets and tweets with images as a percentage of all tweets; and b) all tweets with images as a percentage of all tweets that contained at least one image.}
\label{fig:temporal_overview}
\end{figure}

\descr{Temporal Analysis.} Next, we look into how the tweets from Russian trolls are shared over time with a particular focus on the tweets that contain images.
Figure~\ref{subfig:tweets_temporal} reports the percentage of tweets shared each week normalized by the number of all tweets, while Figure~\ref{subfig:tweets_images_temporal}  the percentages normalized by the number of tweets that contained at least one image.
The former shows that, in the early stages of their operations (before 2016), Russian trolls were posting tweets mostly without images, whereas, after 2016 it seems that they started posting more tweets containing images.
This indicates that they started using more images in their tweets after 2016, likely because they started targeting specific foreign countries (e.g., the US~\cite{muellerreport}), suggesting the Russian trolls might believe the use of images can be better for pushing specific narratives.%

Figure~\ref{subfig:tweets_images_temporal} reveals an overall increase in the use of images after October 2016 with a peak of activity in use of images during the week leading to the Charlottesville rally
in August 2017~\cite{charlotesville}, which led to the death of one counter protester~\cite{charlotesville_death} and was a significant turning point in the use of online hate speech and anti-Semitism in fringe Web communities~\cite{Finkelstein2018}.
This peak likely indicates that the use of images is an effective tactic used by Russian trolls to sow discord on social networks with respect to events related to politics, the alt-right, and white supremacists.

\begin{table}[t]
\centering
\setlength{\tabcolsep}{3pt}
\footnotesize
\begin{tabular}{@{}lrlr@{}}
\toprule
\textbf{Top entity} & \multicolumn{1}{l}{\textbf{\#clusters (\%)}} & \textbf{Top entity} & \multicolumn{1}{l}{\textbf{\#images (\%)}} \\ \midrule
Russia & \multicolumn{1}{r|}{2,783 (3.5\%)} & Russia & 30,426 (4.0\%) \\
Vladimir Putin & \multicolumn{1}{r|}{1,377 (1.7\%)} & Vladimir Putin & 15,718 (2.0\%) \\
Donald Trump & \multicolumn{1}{r|}{1,281 (1.6\%)} & Breaking news & 15,071 (2.0\%) \\
Car & \multicolumn{1}{r|}{1,262 (1.6\%)} & Donald Trump & 13,807 (1.8\%) \\
Ukraine & \multicolumn{1}{r|}{1,031 (1.3\%)} & Car & 10,236 (1.3\%) \\
U.S.A. & \multicolumn{1}{r|}{907 (1.1\%)} & Ukraine & 10,169 (1.3\%) \\
Barack Obama & \multicolumn{1}{r|}{823 (1.0\%)} & U.S.A. & 8,638 (1.1\%) \\
Petro Poroshenko & \multicolumn{1}{r|}{621 (0.8\%)} & Barack Obama & 8,380 (1.1\%) \\
Document & \multicolumn{1}{r|}{530 (0.6\%)} & Petro Poroshenko & 6,654 (0.9\%) \\
Moscow & \multicolumn{1}{r|}{495 (0.6\%)} & Logo & 6.017 (0.8\%) \\
Hillary Clinton & \multicolumn{1}{r|}{479 (0.6\%)} & Moscow & 5,524 (0.7\%) \\
Meme & \multicolumn{1}{r|}{461 (0.6\%)} & Syria & 4,540 (0.6\%) \\
Logo & \multicolumn{1}{r|}{456 (0.6\%)} & Public Relations & 4,459 (0.6\%) \\
Product & \multicolumn{1}{r|}{422 (0.5\%)} & Police & 4,301 (0.6\%) \\
Public Relations & \multicolumn{1}{r|}{416 (0.5\%)} & Hillary Clinton & 4,167 (0.5\%) \\
Illustration & \multicolumn{1}{r|}{393 (0.5\%)} & Document & 4,060 (0.5\%) \\
Syria & \multicolumn{1}{r|}{372 (0.5\%)} & Meme & 3,886 (0.4\%) \\
Web page & \multicolumn{1}{r|}{310 (0.4\%)} & Product & 3,256 (0.4\%) \\
Advertising & \multicolumn{1}{r|}{295 (0.3\%)} & Saint Petersburg & 2,870 (0.4\%) \\
Police & \multicolumn{1}{r|}{290 (0.3\%)} & Illustration & 2,862 (0.4\%) \\ \bottomrule
\end{tabular}%
\caption{Top 20 entities found in images shared by Russian troll accounts.
 We report the top entities both in terms of the number of clusters and of images.}
\label{tbl:top_entities_russians}
\end{table}

\begin{figure*}[t]
\centering
\includegraphics[width=0.665\textwidth]{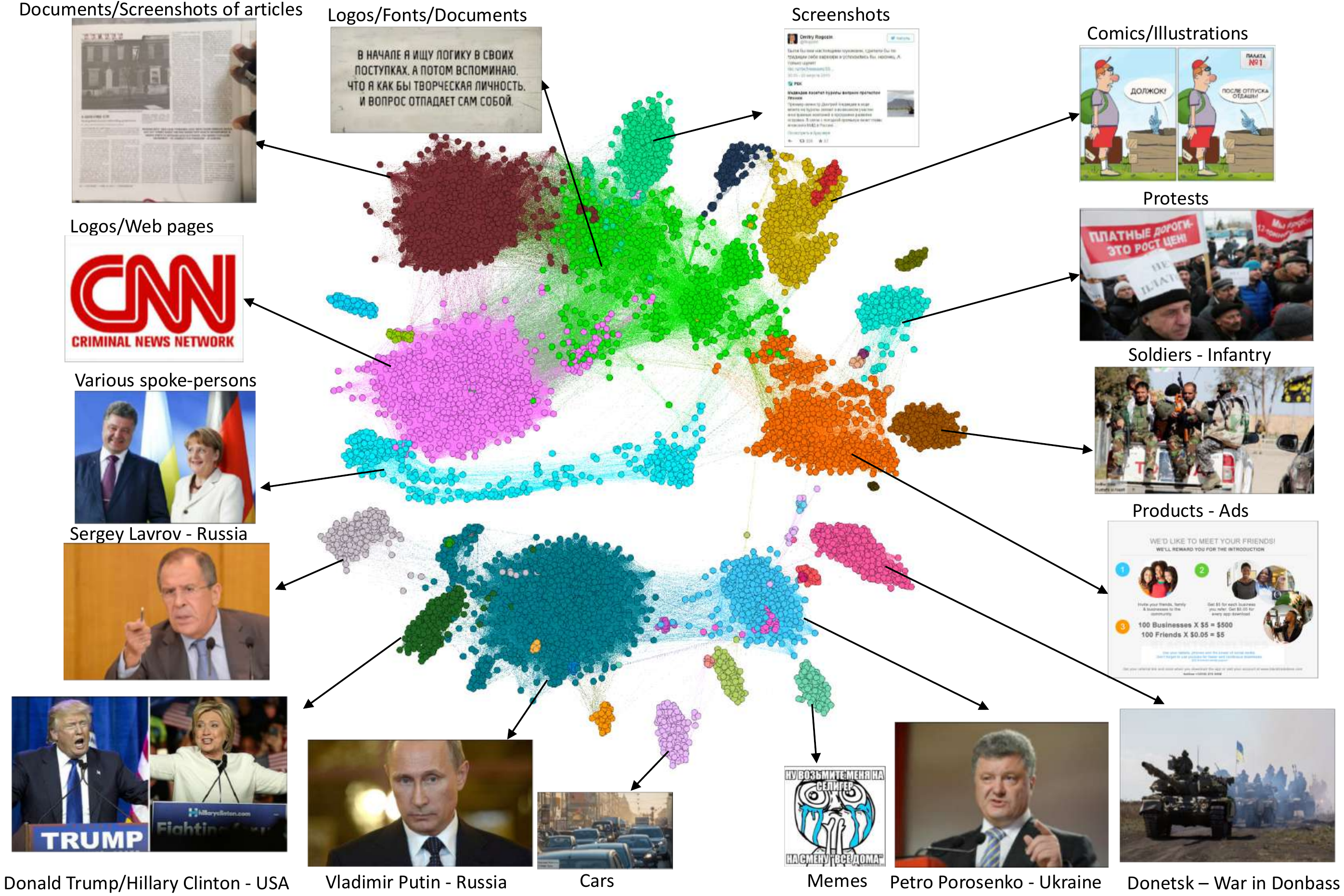}
   \caption{Overview of a subset of the clusters obtained from images shared by the troll accounts.}
\label{fig:russians_clusters_overview}
\end{figure*}

\subsection{Entity Analysis}
We now explore the content of images with a special focus on the \emph{entities} they contain, which allows us to better understand what ``messages'' images were used to convey.
To do so, we use the image processing pipeline presented in~\cite{zannettou2018origins} to create clusters of visually similar images but leverage Google's Cloud Vision API to annotate each cluster (as discussed in the Methodology section).

Then, for each image, we assign the entity with the highest confidence score as returned by the Cloud Vision API.
We also associate the tweet metadata to each image (i.e., which image appears in which tweet).
The final annotated dataset allows us to study the popularity of entities in images posted by state-sponsored accounts on Twitter.

\descr{Popular Entities.} We first look at the popularity of entities for \shortVer{the trolls:}
\shortVer{Table~\ref{tbl:top_entities_russians} reports} the top 20 entities that appear in our image dataset both in terms of the number of clusters, as well as the number of images within the clusters.
We observe that the two most popular entities for Russian trolls are referring to Russia itself (i.e., ``Russia'' and ``Vladimir Putin'' entities).
Also, trolls are mainly focused on events related to Russia, Ukraine, USA, and Syria (their top entities correspond to these countries).
\shortVer{Moreover, several images include screenshots of news articles (see entity ``Web page'') as well as logos of news sites (see entity ``Logo''), hence indicating that these accounts were sharing news articles via images.}
This is because the state-sponsored accounts shared URLs of news articles, which do not include images, hence Twitter automatically adds the logo of the news site to the tweet. %
Finally, we find a non-negligible percentage of images and clusters that show memes, highlighting that memes are exploited by such accounts to disseminate their ideology and probably weaponized information via memes.

\descr{Graph Visualization.} To get a better picture of the spectrum of entities and the interplay between them, we also build \shortVer{a graph, reported in Figure~\ref{fig:russians_clusters_overview},}
where nodes correspond to clusters of images and each edge to the similarity of the entities between the clusters.
For each cluster, we use the set of entities from the Google Cloud Vision API and calculate the Jaccard similarity between each cluster.
Jaccard similarity is useful here, because it exposes meta relationships between clusters.
While images that appear within the same cluster are visually similar, there are likely to be other clusters that represent the same subjects, but from a different visual perspective.
Then, we create an edge between clusters (weighted by their Jaccard similarity) with similarities below a pre-defined threshold.
We set this threshold to 0.4, i.e., we discard all edges between clusters that have a Jaccard similarity less than 0.4, because we want to 1)~capture the main connections between the clusters and 2)~ increase the readability of the graph.
We then perform the following operations:
1)~we run a community detection algorithm using the Louvain method~\cite{blondel2008fast}
and paint each community with a different color;
2)~we lay out the graph with the Force Atlas2 layout~\cite{jacomy2014forceatlas2}, which takes into account weights of edges (i.e., clusters with higher similarity will be positioned closer in the graph);
3)~for readability purposes, we show the top 30\% of nodes according to their degree in the graph; and
4)~we manually annotate the graph with representative images for each community, allowing us to understand the content within each community.
In a nutshell, \shortVer{this graph allows}
us to understand the main communities of entities pushed by the state-sponsored accounts and how they are connected.

\descr{Main Communities.} From Figure~\ref{fig:russians_clusters_overview}, we observe a large community (sapphire) that corresponds to clusters related to Vladimir Putin and Russia.
This community is tightly connected with communities related to Donald Trump/Hillary Clinton/USA (green), Ukraine/Petro Poroshenko (light blue), and Sergey Lavrov (gray).
Also, we observe that other big communities include logos from news outlets (pink) that are tightly connected with communities including screenshots of articles (brown), images of documents (light green), and various other screenshots (emerald).
Other communities worth noting are those including comics and various illustrations (yellow) as well as images of products and advertisements (orange).
Overall, these findings highlight that state-sponsored troll accounts shared many images with a wide variety of themes, ranging from memes to news via screenshots.

\begin{table}[]
\centering
\resizebox{\columnwidth}{!}{%
\begin{tabular}{@{}lrlr@{}}
\toprule
\textbf{Domain} & \multicolumn{1}{l}{\textbf{\#clusters (\%)}} & \textbf{Domain} & \multicolumn{1}{l}{\textbf{\#images (\%)}} \\ \midrule
pinterest.com & \multicolumn{1}{r|}{9,433 (12.0\%)} & pinterest.com & 76,231 (10.1\%) \\
twitter.com & \multicolumn{1}{r|}{5,481 (7.0\%)} & twitter.com & 46,609 (6.1\%) \\
youtube.com & \multicolumn{1}{r|}{4,132 (5.2\%)} & youtube.com & 40,540 (5.4\%) \\
wordpress.com & \multicolumn{1}{r|}{3,329 (4.2\%)} & riafan.ru & 35,497 (4.7\%) \\
ria.ru & \multicolumn{1}{r|}{3,260 (4.1\%)} & ria.ru & 31.153 (4.1\%) \\
riafan.ru & \multicolumn{1}{r|}{2,734 (3.4\%)} & wordpress.com & 30,464 (4.0\%) \\
blogspot.com & \multicolumn{1}{r|}{2,432 (3.0\%)} & blogspot.com & 20,890 (2.7\%) \\
livejournal.com & \multicolumn{1}{r|}{2,381 (3.0\%)} & sputniknews.com & 20,558 (2.7\%) \\
pikabu.ru & \multicolumn{1}{r|}{2,073 (2.6\%)} & livejournal.com & 20,227 (2.6\%) \\
me.me & \multicolumn{1}{r|}{1,984 (2.5\%)} & pikabu.ru & 17,250 (2.2\%) \\
sputniknews.com & \multicolumn{1}{r|}{1,943 (2.4\%)} & rambler.ru & 15,227 (2.0\%) \\
reddit.com & \multicolumn{1}{r|}{1,826 (2.3\%)} & me.me & 14,675 (1.9\%) \\
theguardian.com & \multicolumn{1}{r|}{1,527 (1.9\%)} & theguardian.com & 14,111 (1.9\%) \\
rambler.ru & \multicolumn{1}{r|}{1,524 (1.9\%)} & reddit.com & 14,025 (1.8\%) \\
facebook.com & \multicolumn{1}{r|}{1,336 (1.7\%)} & wikimedia.org & 12,897 (1.7\%) \\
dailymail.co.uk & \multicolumn{1}{r|}{1,271 (1.6\%)} & wikipedia.org & 12,081 (1.6\%) \\
imgur.com & \multicolumn{1}{r|}{1,210 (1.5\%)} & facebook.com & 12,012 (1.6\%) \\
wikimedia.org & \multicolumn{1}{r|}{1,051 (1.3\%)} & dailymail.co.uk & 9,854 (1.3\%) \\
pinterest.co.uk & \multicolumn{1}{r|}{1,027 (1.3\%)} & imgur.com & 9,381 (1.2\%) \\
wikipedia.org & \multicolumn{1}{r|}{996 (1.2\%)} & cnn.com & 8,606 (1.1\%) \\ \bottomrule
\end{tabular}%
}
\caption{Top 20 domains that shared the same images as the trolls. We report the top domains both in terms of number of clusters and number of images within the clusters.}
\label{tbl:top_domains_russians}
\end{table}

\begin{figure*}[t]
\centering
\includegraphics[width=0.6\textwidth]{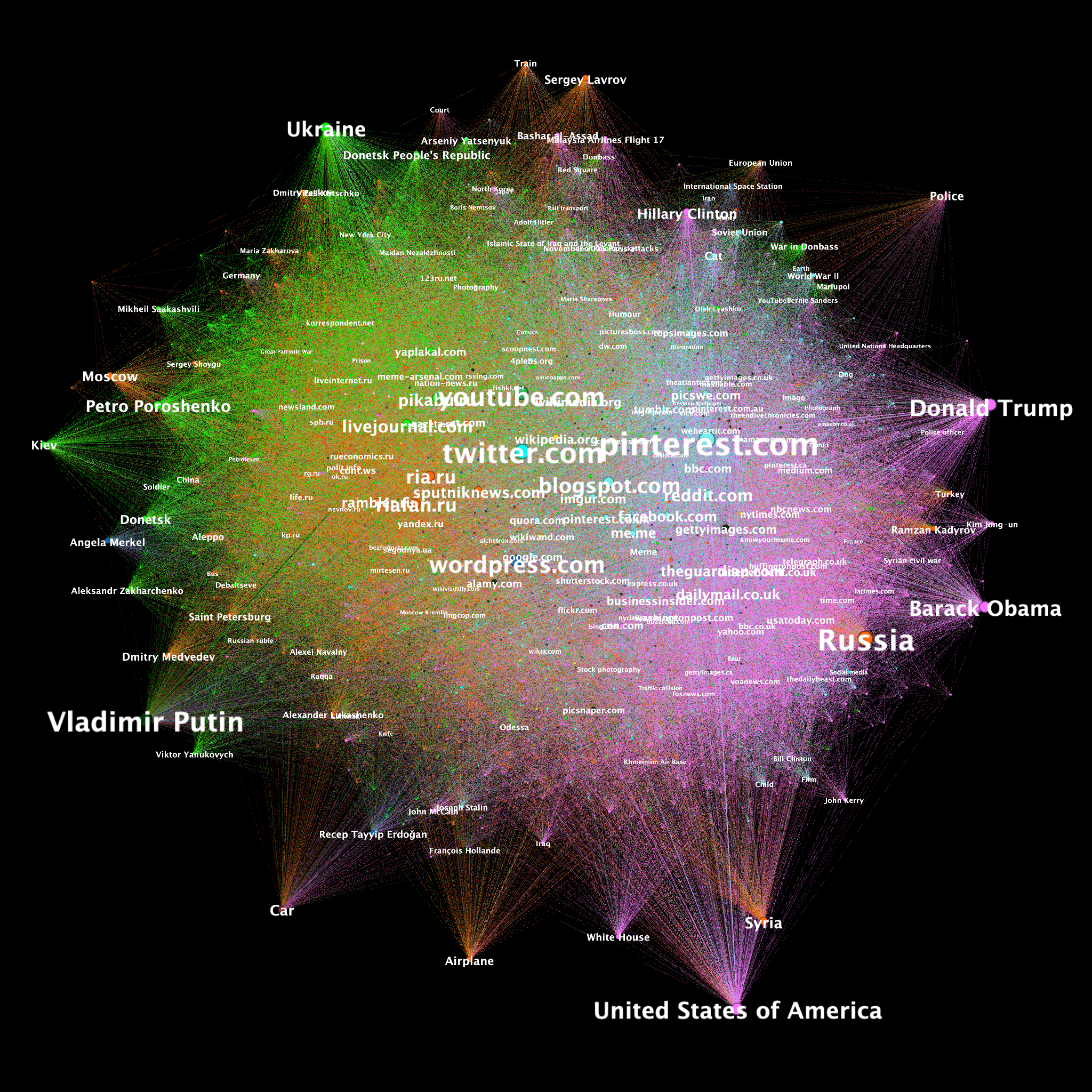}
   \caption{Visualization of the interplay between entities and domains that shared matching images as the ones shared by the trolls.}
   \label{fig:russians_entities_domains_graph}
\end{figure*}

\subsection{Images Occurrence across the Web}\label{sec:greater}
Our next set of measurements analyze the co-occurrence of the images posted by Russian state-sponsored accounts across the greater Web. %
Recall that the Cloud Vision API also provides details about the appearance of an image across the Web.
This is useful when studying the behavior of state-sponsored accounts, as it either denotes that they posted the images on other domains too, or they obtained the image from a different domain, or that other users on the Web posted them on other domains too.
Thus, studying the domains that shared the same images as state-sponsored accounts allows us to understand their behavior and potential impact on the greater Web.
For instance, this information can be used to detect domains that are exclusively controlled by state-sponsored actors to spread disinformation.

\shortVer{In Table~\ref{tbl:top_domains_russians}, we} report the top domains, both in terms of number of clusters and number images within the clusters, that shared the same images as the state-sponsored accounts\shortVer{.}
Unsurprisingly, the most popular domains are actually mainstream social networking sites (e.g., Pinterest, Twitter, YouTube, and Facebook)\shortVer{.}
Also, among the popular domains we find popular Russian news outlets like \url{ria.ru} and \url{riafan.ru}, as well as Russian-owned social networking sites like \url{livejournal.com}\footnote{Although founded in the US, LiveJournal was sold to a Russian company in 2007, and all servers have been located in Russia since 2017.} and \url{pikabu.ru}.
This highlights the efforts by Russian trolls to sway public opinion about public matters related to Russia.
We further find both mainstream and alternative news outlets like \url{theguardian.com} and \url{sputniknews.com}, respectively (we use the list provided by~\cite{zannettou2017web} to distinguish mainstream and alternative news outlets).
This provides evidence that the efforts of Russian trolls had an impact on, or were inspired by, content shared on a wide variety of important sites in the information ecosystem on the Web.

Next, we aim to provide a holistic view of the domains while considering the interplay between the entities of the images and the domains that they also shared them.
To do this, we create \shortVer{a graph} where nodes are either entities or domains that were returned from the Cloud Vision API.
An edge exists between a domain node and an entity node if an image appearing on the domain contained the given entity.
Then, we perform the operations (1) and (2) as described in the entities analysis section (i.e., community detection and layout algorithm).
\shortVer{We do this for the images posted by the trolls and present the resulting graph in Figure~\ref{fig:russians_entities_domains_graph}.}
\shortVer{This graph allows} us to understand which domains shared images pertaining to various semantic entities.
We find popular Web communities like Twitter, Pinterest, Facebook and YouTube in the middle of the graph, constituting a separate community (light blue), i.e., they are used for sharing images across all entities.
Entities mainly related to Russia are shared via Russian state-sponsored outlets like \url{sputniknews.com} (see orange community).
Entities that are related to the USA and political persons like Donald Trump, Barack Obama, and Hillary Clinton are part of a separate community (pink) with popular news outlets like \url{washingtonpost.com} and \url{nytimes.com}.
Finally, for matters related to Ukraine (green community) most of the images co-appeared on popular Russian-owned social networks like \url{livejournal.com} and \url{pikabu.ru}.

Overall, our findings indicate that the same images often appear on both their feeds and specific domains.
Thus, state-sponsored trolls might be trying to make their accounts look more credible and push their agenda by targeting unwitting users on popular Web communities like Twitter.

\section{Cross-Platform Influence}

Our analysis above studies the occurrence of images shared by Russian state-sponsored accounts on other domains, but does not encapsulate the interplay between multiple communities.
In reality, the Web consists of a large number of communities that are not exclusively independent of each other, as communities naturally influence each other.
For instance, a Twitter user might share an image on Twitter that she previously saw on Reddit: in this case, we see that the Reddit community has ``influenced'' the Twitter community with respect to the sharing of that particular image.

In this section, we model and measure the interplay and the influence across Web communities with respect to the dissemination of images that were also shared by Russian state-sponsored accounts on Twitter.
In other words, we aim to understand how {\em influential} the trolls were in spreading images to other communities.
To do so, we rely on Hawkes Processes~\cite{linderman2014,lindermanArxiv}, which allow us to estimate the probabilities that an appearance of an image on one community is due to that image previously occurring on the same or on another Web community.
\vspace{0.2cm}

\subsection{Hawkes Processes}

Hawkes Processes are self-exciting temporal point processes~\cite{hawkes1971spectra} that describe how {\em events} occur on a set of {\em processes.}
In our setting, events are the posting of an image, and processes are Web communities.
Generally, a Hawkes model consists of $K$ processes; each process has a \emph{rate} of events that dictates the frequency of the creation of events in the specific process.
The occurrence of an event on one process, causes \emph{impulses} to the rest of the processes, temporarily increasing the rate of events in the other processes.
The impulses comprise two useful pieces of information: the intensity of the increase in the rate, and how it is distributed and decays over time.

By fitting a Hawkes model using Bayesian inference to data that describes the appearances of events in the processes, we obtain the parameter values for the impulses.
This lets us quantify the overall rate of events in each process, as well as how much previous events contribute to the rate, at a given point in time. %
Naturally, we cannot possibly know what exactly triggered the creation of an event on a process, %
however, we can use Hawkes Processes to calculate the {\em probability} that the cause of an event is another process in the model, as also done by previous work~\cite{zannettou2017web,zannettou2018origins}.
Note that the background rate of the Hawkes Processes allow us to capture and model the interplay of external sources (i.e., platforms that we do not use in our analysis), hence the resulting probabilities encapsulate the influence of the greater Web via the background rates.

\begin{table*}[]
\centering
\footnotesize
\begin{tabular}{@{}lrrrrrrrr@{}}
\toprule
\textbf{}                                & \multicolumn{1}{l}{\textbf{/pol/}} & \multicolumn{1}{l}{\textbf{Reddit}} & \multicolumn{1}{l}{\textbf{Twitter}} & \multicolumn{1}{l}{\textbf{Gab}} & \multicolumn{1}{l}{\textbf{T\_D}} & \multicolumn{1}{l}{\textbf{Russia}} & \multicolumn{1}{l}{\textbf{Total Events}} & \multicolumn{1}{l}{\textbf{pHashes}} \\ \midrule
\textbf{Republican Party-related Images} & 96,569                             & 85,457                              & 145,372                              & 18,496                           & 21,733                            & \multicolumn{1}{r|}{18,332}         & \multicolumn{1}{r|}{385,959}              & 9,947                                \\
\textbf{Democratic Party-related Images} & 64,282                             & 38,602                              & 96,082                               & 13,485                           & 17,797                            & \multicolumn{1}{r|}{12,465}         & \multicolumn{1}{r|}{242,713}              & 6,043                                \\
\textbf{All images}                      & 409,026                            & 421,115                             & 1,904,570                            & 75,361                           & 72,679                            & \multicolumn{1}{r|}{231,730}        & \multicolumn{1}{r|}{3,114,481}            & 90,299                               \\ \bottomrule
\end{tabular}%
\caption{Number of events (image occurrences) for images shared by Russian state-sponsored accounts. We report the number of events on Twitter (other users), Russian state-sponsored accounts on Twitter (Russia), Gab, /pol/, Reddit, and The\_Donald subreddit.}
\label{tbl:hawkes_events}
\end{table*}

\subsection{Datasets}

\descr{Cross-Platform Dataset.} We use a publicly available dataset consisting of 160M pHashes and image URLs for all the images posted on Twitter (using 1\% Streaming API), Reddit, 4chan's \dspol, and Gab, between July 2016 and July 2017.\footnote{\url{https://zenodo.org/record/1451841}.}
Then, we %
select the images that have the same pHashes with the ones shared by Russian state-sponsored accounts on Twitter.
For each one of these images, we find all their occurrences on Reddit, \dspol, Gab, and Twitter.
Next, we omit images that appear less than five times across all Web communities we study, ultimately obtaining a set of 90K pHashes.
Finally, we annotate each pHash using the Web entities obtained from the Cloud Vision API.

Since our dataset focuses primarily on the year before and after the 2016 US elections, we concentrate our analysis around this major event.
Specifically, we want to assess whether Russian state-sponsored accounts were more effective in pushing images related to the Democratic Party or Republican Party.
To do so, we select all the pHashes that have a Cloud Vision Web entity corresponding to ``Democratic Party,'' ``Hillary Clinton,'' and ``Barack Obama'' for the Democratic Party, and ``Republican Party'' and ``Donald Trump'' for the Republican Party.
Using these entities, we find 9.9K images related to the Republican Party and 6K images related to the Democratic Party.

\begin{figure}[t]
\centering
\includegraphics[width=\columnwidth]{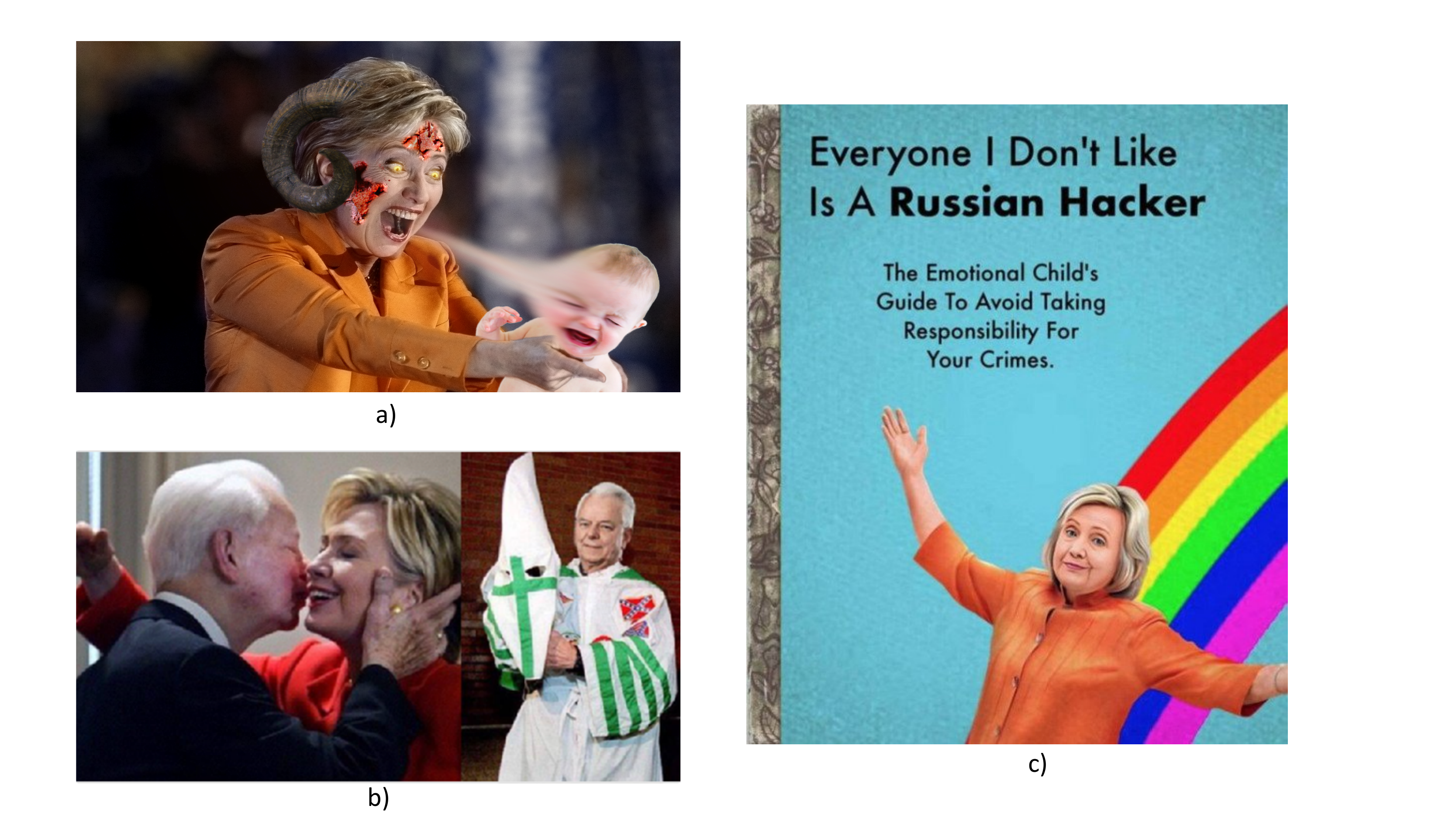}
   \caption{Examples of images in the Democratic Party sample.}
\label{fig:dems_examples}
\end{figure}

\begin{figure}[t]
\centering
\includegraphics[width=0.87\columnwidth]{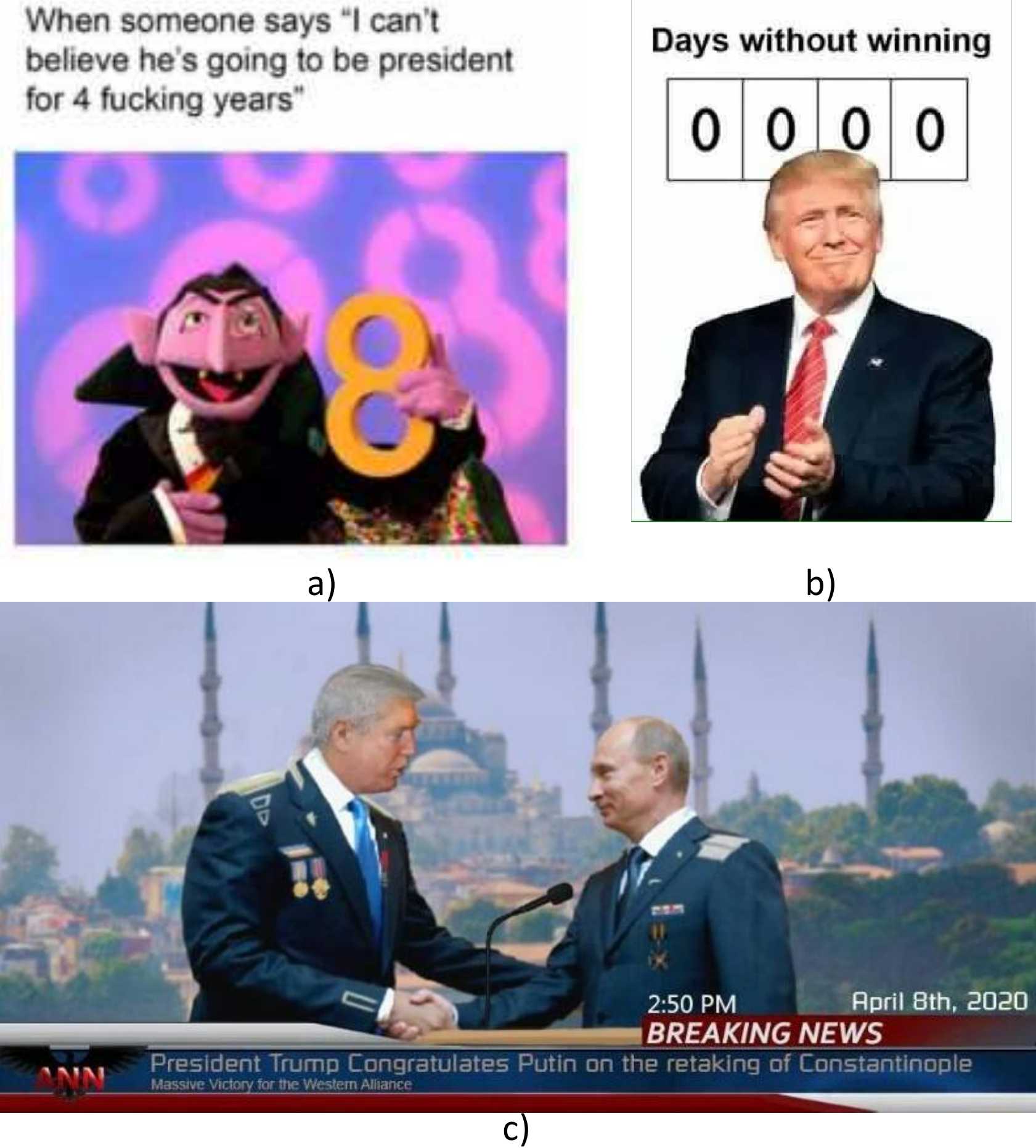}
   \caption{Examples of images  in the Republicans Party sample.}
\label{fig:reps_examples}
\end{figure}

\descr{Examples.} To provide an intuition on what some of the politically charged images look like, we provide some examples in Figure~\ref{fig:dems_examples} and Figure~\ref{fig:reps_examples} for the Democratic and the Republican party, respectively.
These illustrate how Russian trolls use images to spread disinformation: for instance, Figure~\ref{fig:dems_examples}(b) shows Senator Robert Byrd meeting with Hillary Clinton and, in another occasion, wearing a Ku Klux Klan robe.
The image with the robe is known to be fake as reported later by Snopes~\cite{snopes_kkk}.
We can also observe how state-sponsored accounts rely on edited/photoshopped images to make specific personalities look bad: e.g., Figure~\ref{fig:dems_examples}(a) is an edited image aimed at reinforcing  the idea that Hillary Clinton was involved in the Pizzagate conspiracy theory (her face was edited and a baby was added in the picture).
Finally, we find several memes that are meant to be funny; however they have a strong political nature and can effectively disseminate ideology.
For instance, Figure~\ref{fig:dems_examples}(c) makes fun of Hillary Clinton, while Figure~\ref{fig:reps_examples}(a) and Figure~\ref{fig:reps_examples}(b) are clearly pro-Trump and celebrate him winning the 2016 US elections.

\begin{figure*}[t!]
\center
\subfigure[]{
\includegraphics[width=0.45\textwidth]{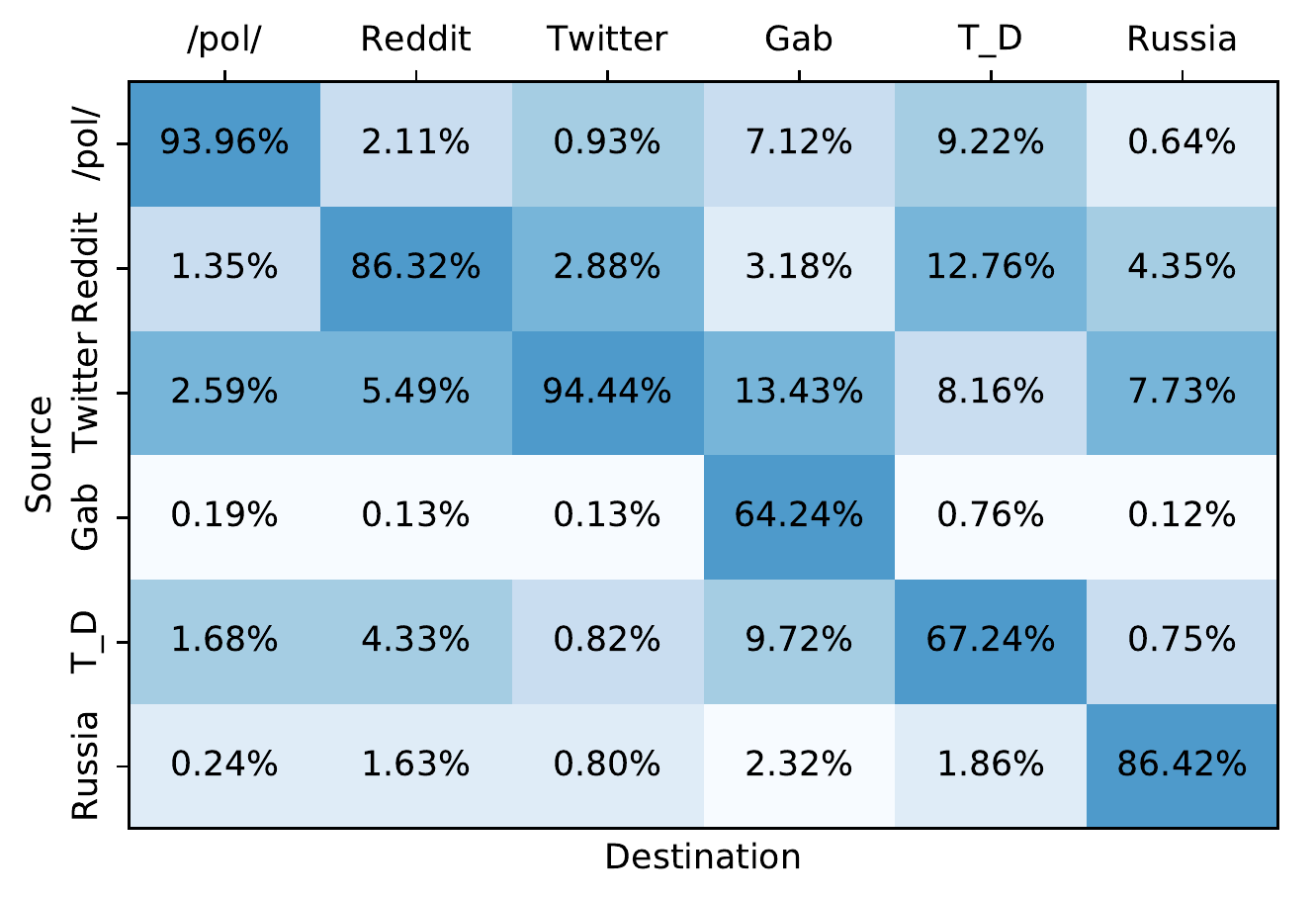}\label{subfig:hawkes_all_influence}}
\subfigure[]{
\includegraphics[width=0.45\textwidth]{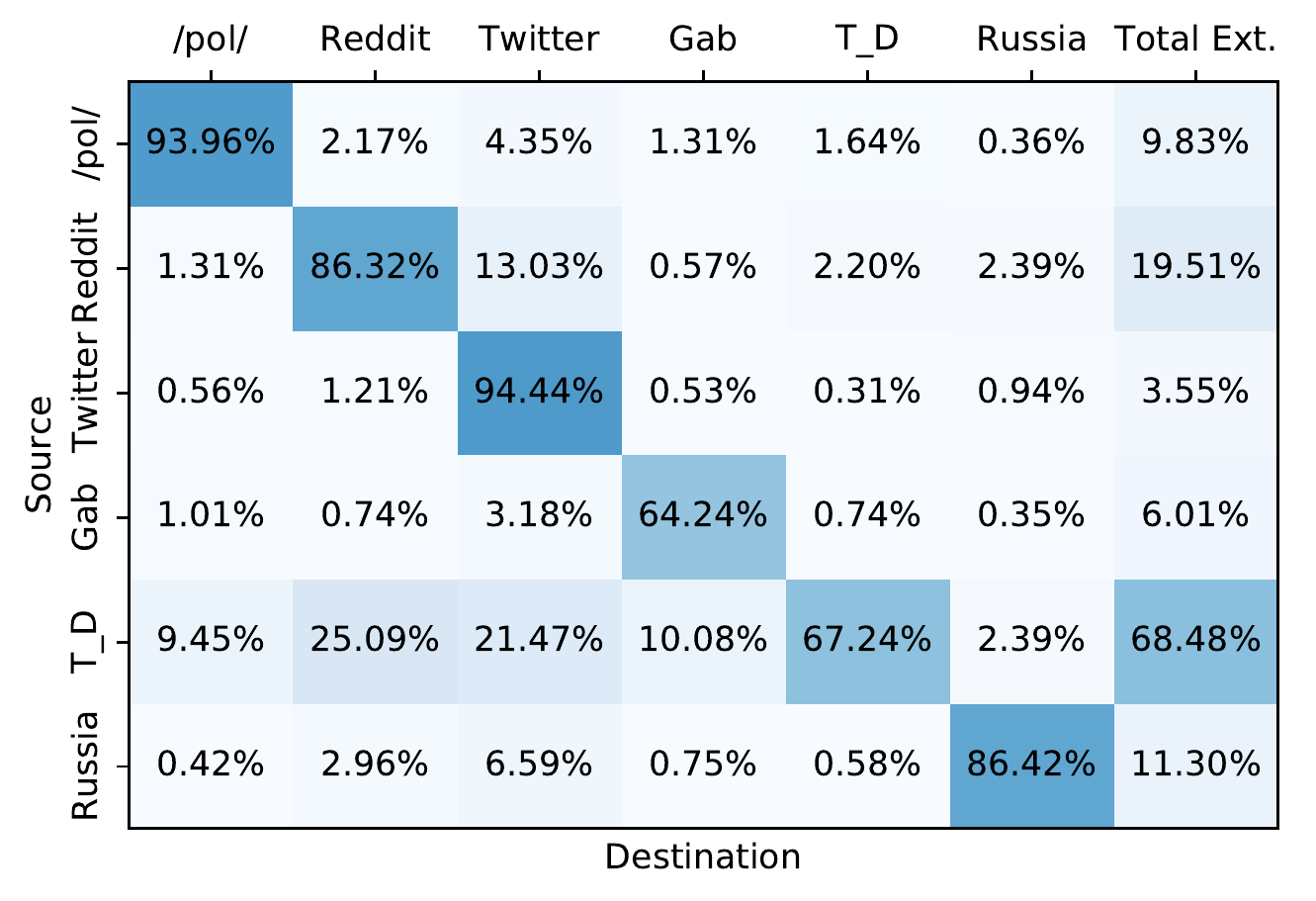}\label{subfig:hawkes_all_normalized}}
  \caption{Influence estimation for all images shared by Russian state-sponsored accounts on Twitter: a) Raw {\em influence} between source and destination Web communities; and b) Normalized influence ({\em efficiency}) of each Web community as the results are normalized by the numbers of events created on the source community. The numbers in the cell can be interpreted as the expected percentage of events created on the destination community because of previously occurring events on the source community.}
\label{fig:hawkes_all}
\end{figure*}

\descr{Events.} Table~\ref{tbl:hawkes_events} summarizes the number of events for our dataset.
Note that we elect to decouple The\_Donald subreddit from the rest of Reddit mainly because of its strong political nature and support towards Donald Trump~\cite{flores2018mobilizing}. %
By looking at the raw numbers of events per category, we note that in general Russian state-sponsored accounts shared more content related to the Republican Party when compared to the Democratic Party.
The same applies for all the other communities we study: in general we find 1.59 times more events for the Republican Party than the Democratic Party (385K vs 242K events).
This indicates that content related to the Republican Party was more popular in all Web communities during this time period and that Russian state-sponsored accounts pushed more content related to the Republican Party, likely in favor of Donald Trump as previous research show~\cite{zannettou2018let}.

\subsection{Results}
We create a Hawkes model for each pHash. Each model consists of six processes, one for each of Reddit, The\_Donald subreddit, Gab, Russian state-sponsored accounts on Twitter, and other Twitter users.
Then, we fit a Hawkes model using Gibbs sampling as described in~\cite{linderman2014} for each of the 90K pHashes.

\descr{Metrics.} After fitting the models and obtaining all the parameters for the models, following the methodology presented in~\cite{zannettou2018origins}, we calculate the \emph{influence} and \emph{efficiency} that each community had to each other.
The former denotes the percentage of events (i.e., image appearances) on a specific community that appear because of previous events on another community, while the latter is a normalized influence metric that denotes how efficient a community is in spreading images to the other communities irrespectively to the number of events that are created within the community.
In other words, efficiency describes how influential the posting of a single event to a particular community is, with respect to how it spreads to the other communities.

\begin{figure*}[t!]
\center
\subfigure[]{
\includegraphics[width=0.45\textwidth]{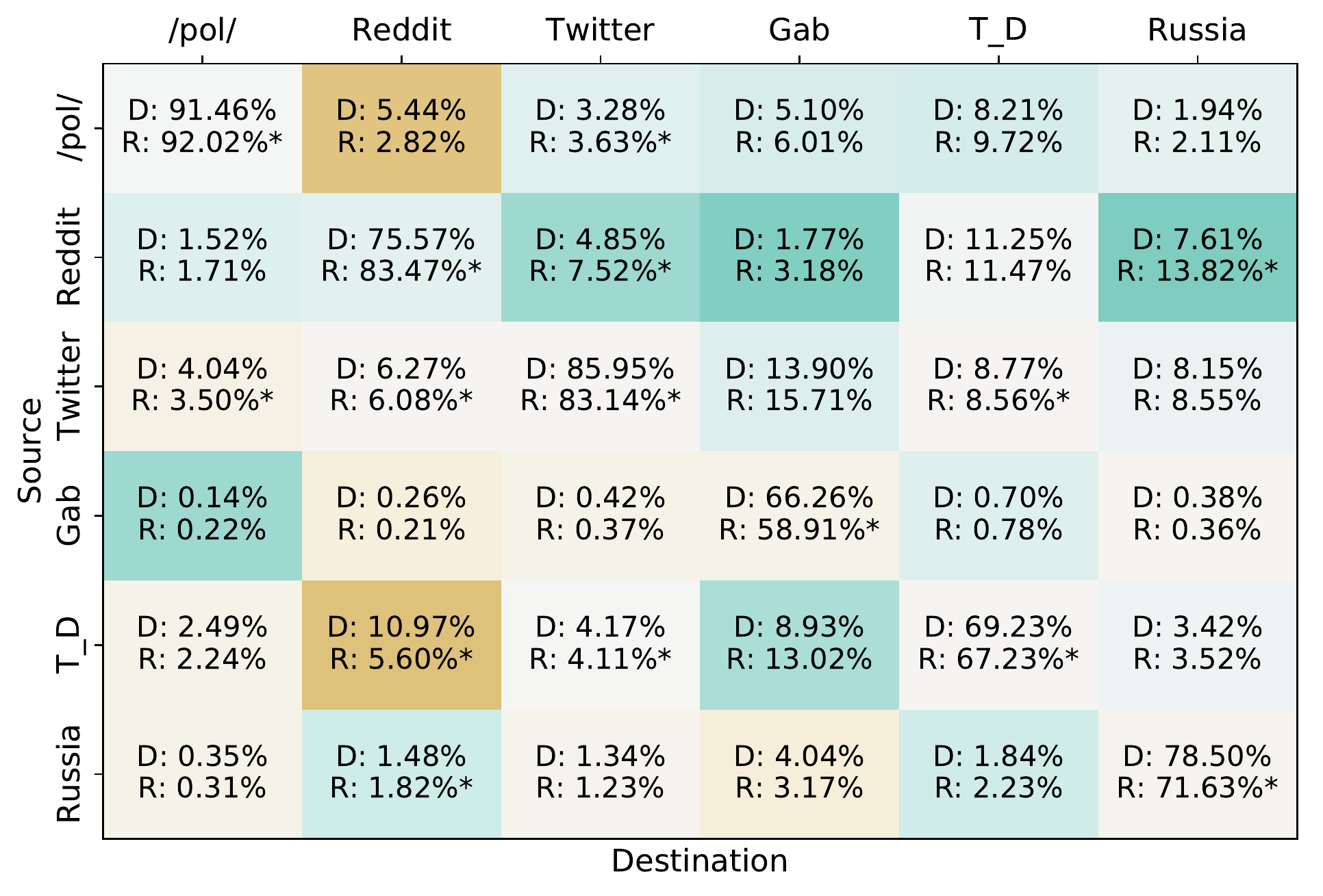}\label{fig:hawkes_all_influence_democrats_republicans}}
\subfigure[]{
\includegraphics[width=0.45\textwidth]{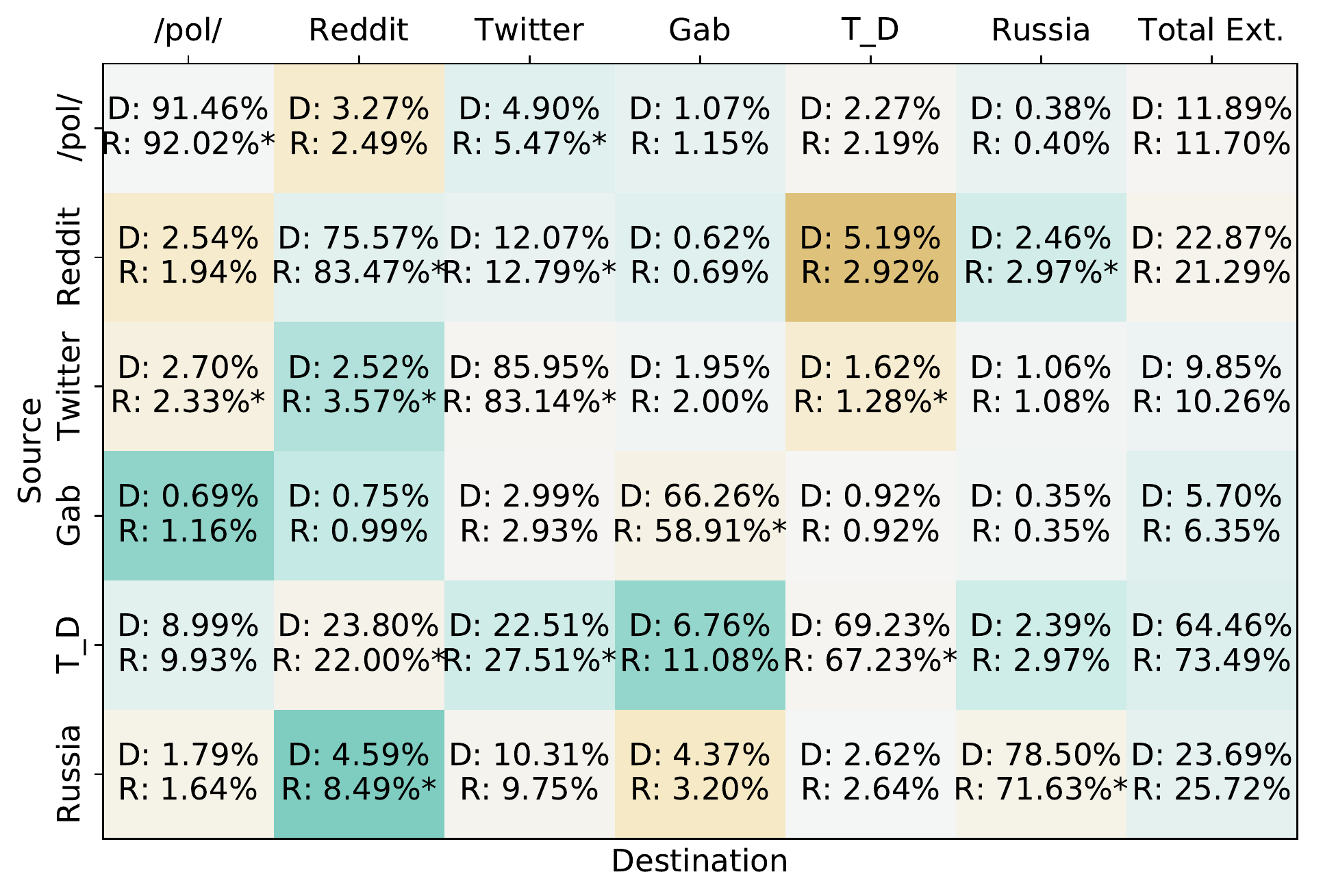}\label{fig:hawkes_all_normalized_democrats_republicans}}
  \caption{Influence estimation for images shared by Russian state-sponsored accounts on Twitter related to the Republican party (R) and the Democratic Party (D): a) Raw {\em influence} between source and destination Web communities; and b) Normalized influence ({\em efficiency}) of each Web community as the results are normalized by the numbers of events created on the source community. }
\label{fig:hawkes_democrats_republicans}
\end{figure*}

\descr{Overall Influence \& Efficiency.}
Figure~\ref{fig:hawkes_all} reports the influence estimation results for all the events (i.e., all the images that were shared by Russian state-sponsored accounts and have at least five occurrences across all the Web communities we study).
When looking at the raw influence results (Figure~\ref{subfig:hawkes_all_influence}), we observe that Russian state-sponsored accounts had the most influence towards Gab (2.3\%), followed by The\_Donald subreddit (1.8\%), and the rest of Reddit (1.6\%), while they had the least influence to 4chan's \dspol (0.2\%).
By comparing the influence of regular Twitter users, with respect to the dissemination of memes, to the influence of the state-sponsored actors (see Figure 11 in extended version of~\cite{zannettou2018origins}\footnote{Available via \url{https://arxiv.org/abs/1805.12512}}), we observe similar patterns.
That is, regular Twitter users were more influential on Gab (8\%), followed by The\_Donald (3.6\%), and the rest of Reddit (2.8\%), while they had the least influence on \dspol (0.7\%).
This comparison indicates that Russian trolls influenced other platforms similarly to regular Twitter users with the difference that the intensity of their influence is substantially lower (between 3.5x-1.5x times lower), mainly due to the fact that Russian trolls consist of a few thousands accounts.
Furthermore, when comparing the results for Twitter against previous characterizations of Russian trolls on news URLs (see Figure 14 (a) in ~\cite{zannettou2018let}), we find that actually Russian trolls were more influential in spreading news URLs compared to images (1.29\% for news URLs and 0.8\% for images).

When looking at the efficiency of Russian state-sponsored accounts (last row in Figure~\ref{subfig:hawkes_all_normalized}), we find that they were most efficient in pushing the images on Twitter (6.5\%) likely because it is the same social network.
Also, they were particularly efficient in pushing images towards the rest of Reddit (2.9\%), while again we find that they were not very effective towards 4chan's \dspol (0.4\%).
Furthermore, we report the overall external efficiency of each community towards all the other communities (right-most column in Figure~\ref{subfig:hawkes_all_normalized}).
We find that the most efficient platform in the ones that we study is The\_Donald subreddit (68.4\%), followed by the rest of Reddit (19.5\%) and the Russian state-sponsored accounts on Twitter (11.3\%).
Again, by looking at previous results based on news (see Figure 15 (a) in ~\cite{zannettou2018let}), we observe that Russian trolls were more efficient in spreading news URLs compared to images (16.95\% external influence for news, while for images we find 11.3\%).

\descr{Politics-related Images.}
Next, we investigate how our influence estimation results change when considering only the politics-related images, and in particular the differences between the images pertaining to the Republican and Democratic Parties.
Figure~\ref{fig:hawkes_democrats_republicans} reports our influence and efficiency estimation results for the images related to the Republican Party (R) and Democratic Party (D).
{\em NB:} To assess the statistical significance of these results, we perform a two-sample Kolmogorov-Smirnov test to the influence distributions of the two samples and annotate the figures with an * for cases where $p<0.01$.
We make the following observations.
First, Russian state-sponsored accounts were most influential in pushing both Democratic and Republican Party-related images to Gab, The\_Donald subreddit, and the rest of the Reddit, while again were the least influential in spreading these images in 4chan's \dspol (see last row in Figure~\ref{fig:hawkes_all_influence_democrats_republicans}).

Second, when comparing the results for both parties, we observe that on Twitter they have more or less the same influence for both Republicans and Democratic parties (1.3\% vs 1.2\%), on Gab they were more influential in spreading Democratic Party images when compared to Republican party (4.0\% vs 3.1\%).
For The\_Donald and the rest of Reddit we observe the opposite: they were more influential in spreading Republican Party related images when compared to the Democratic Party (see last row in Figure~\ref{fig:hawkes_all_influence_democrats_republicans}).

Third, by looking at the efficiency results (Figure~\ref{fig:hawkes_all_normalized_democrats_republicans}), we find that again that Russian state-sponsored accounts were most efficient in spreading political images to big mainstream communities like Twitter and Reddit (see last row in Figure~\ref{fig:hawkes_all_normalized_democrats_republicans}).
Fourth, by looking at the overall external influence of the communities (right-most column in Figure~\ref{fig:hawkes_all_normalized_democrats_republicans}), we observe that again The\_Donald subreddit had the bigger efficiency (over 60\% for both parties), followed by the Russian state-sponsored accounts on Twitter and the rest of Reddit.
Finally, by comparing the efficiency of state-sponsored trolls on all images vs the political-related images (cf. Figure~\ref{subfig:hawkes_all_normalized} and Figure~\ref{fig:hawkes_all_normalized_democrats_republicans}), we find that Russian state-sponsored trolls were over 2 times more efficient in spreading political-related imagery when compared to all the images in our dataset (11.3\% vs 23.6\% and 25.7\%).

\begin{figure*}[t!]
\center
\subfigure[]{
\includegraphics[width=0.28\textwidth]{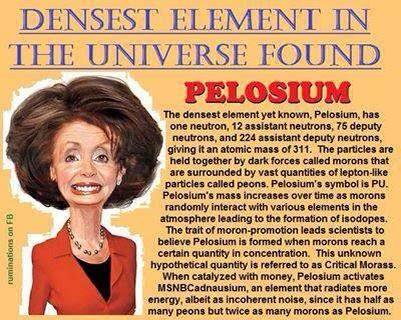}\label{fig:most_influential_dems_1}}
~~~
\subfigure[]{
\includegraphics[width=0.28\textwidth]{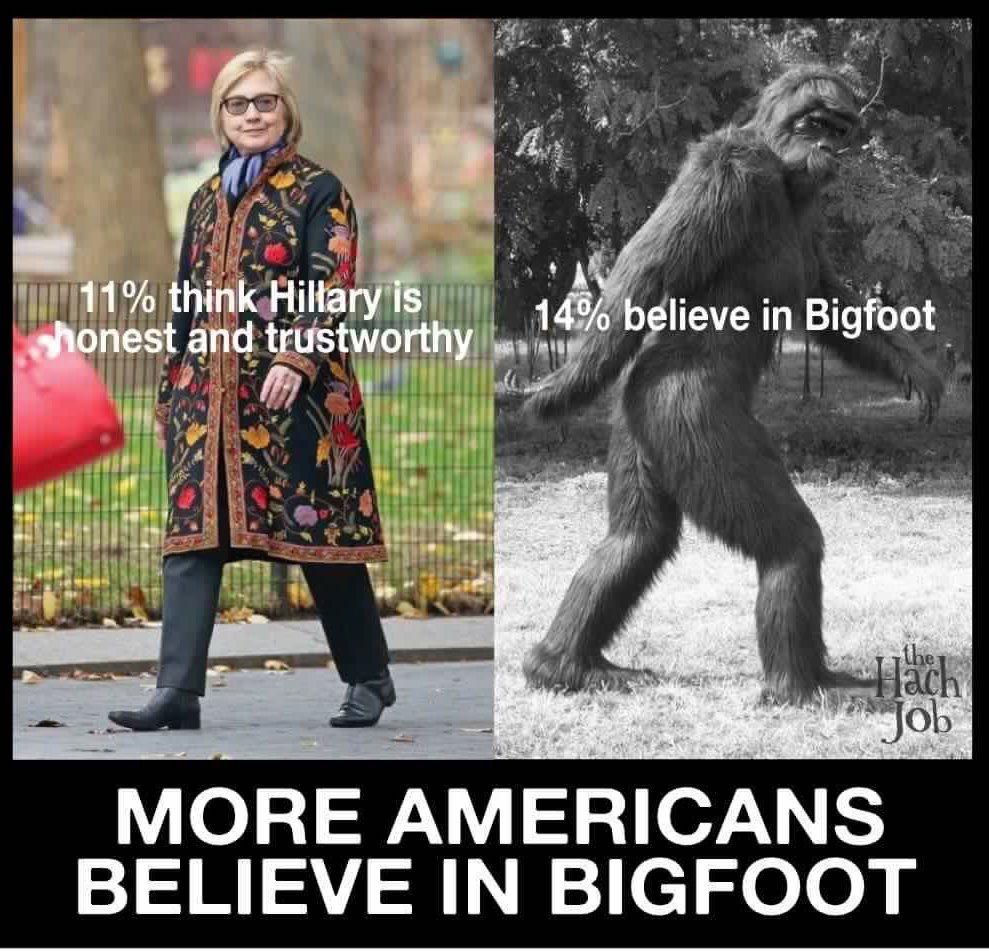}\label{fig:most_influential_dems_2}}
~~~
\subfigure[]{
\includegraphics[width=0.28\textwidth]{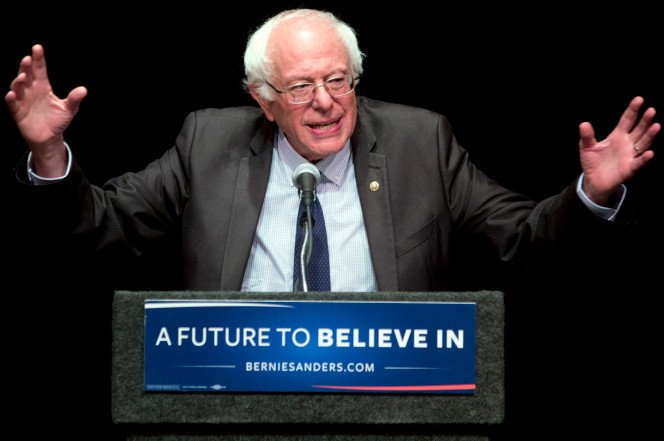}\label{fig:most_influential_dems_3}}
  \caption{Top three most influential images related to the Democratic Party shared by Russian state-sponsored accounts.}
  \label{fig:most_influential_images_dems}
\end{figure*}

\begin{figure*}[t!]
\center
\subfigure[]{
\includegraphics[width=0.28\textwidth]{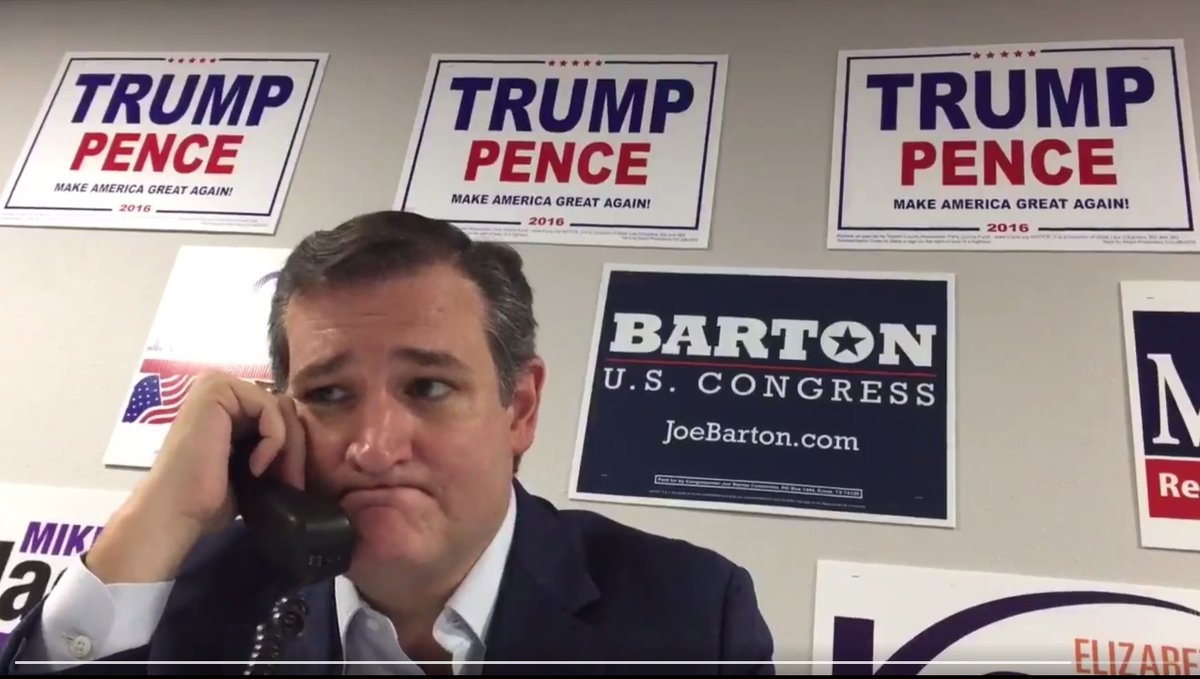}\label{fig:most_influential_reps_1}}
~~~
\subfigure[]{
\includegraphics[width=0.28\textwidth]{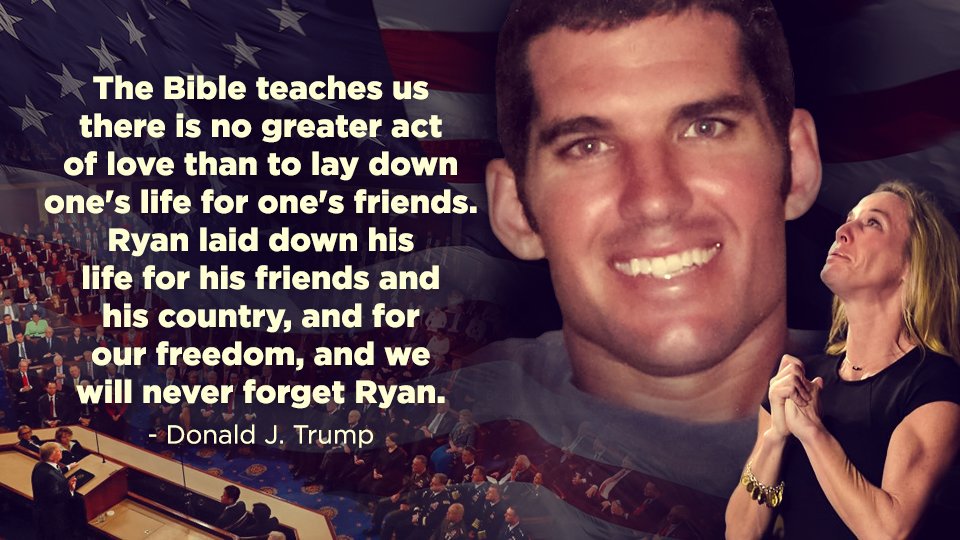}\label{fig:most_influential_reps_2}}
~~~
\subfigure[]{
\includegraphics[width=0.28\textwidth]{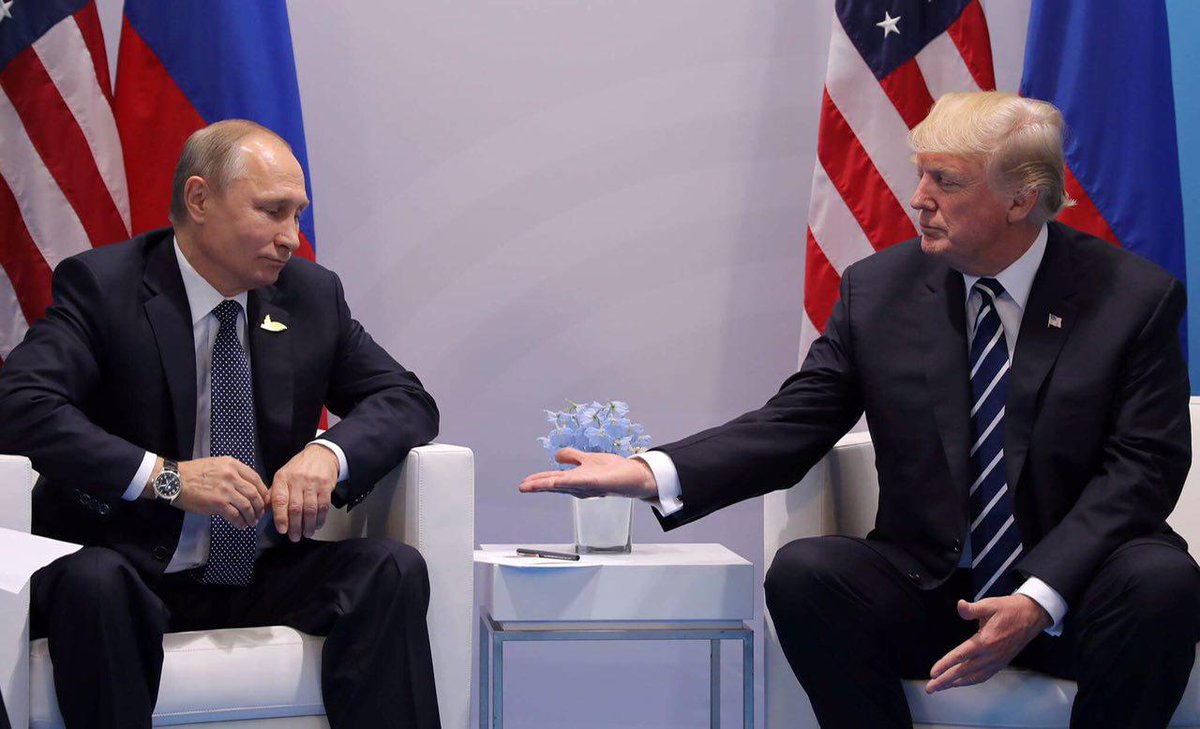}\label{fig:most_influential_reps_3}}
  \caption{Top three most influential images related to the Republican Party shared by Russian state-sponsored accounts.}
  \label{fig:most_influential_images_reps}
\end{figure*}

\descr{Most Influential Images.} Since our influence estimation experiments are done with the granularity of specific pHashes, we can also assess {\em which} images the Russian state-sponsored accounts were more influential in spreading.
To do so, we sort the influence results for the Democratic and Republican Parties according to the external influence that Russian state-sponsored accounts had to all the other Web communities, and report the top three images with the most influence.
Figure~\ref{fig:most_influential_images_dems} and Figure~\ref{fig:most_influential_images_reps} show the three most influential images shared by Russian state-sponsored accounts for the Democratic and Republicans party, respectively.

Evidently, Russian state-sponsored accounts were particularly influential in spreading images ``against'' the Democratic Party: for instance, Figure~\ref{fig:most_influential_dems_1} is an image that trolls Nancy Pelosi, currently serving as speaker of the US House of Representatives, while Figure~\ref{fig:most_influential_dems_2} shares a political message against Hillary Clinton's chances during the 2016 US elections.
On the other hand, the most influential images related to the Republican Party (Figure~\ref{fig:most_influential_images_reps}) are neutral and likely aim to disseminate pro-Trump messages and imagery.

\section{Discussion \& Conclusion} \label{sec:conclusion}
This paper presented a large-scale quantitative analysis of 1.8M images shared by Russian state-sponsored accounts (``Russian trolls'') on Twitter.
Our work is motivated, among other things, by the fact that social network users tend to put little effort into verifying information and they are often driven by visual cues, e.g., images, for re-sharing content~\cite{gabielkov2016social}.
Therefore, as state-sponsored accounts tend to post disinformation~\cite{mejias2017disinformation}, analyzing the images they share represents a crucial step toward understanding and countering the spread of false information on the Web, and its impact on society.

By extending the image processing pipeline presented in~\cite{zannettou2018origins}, we clustered the images and annotated them using Google's Cloud Vision API.
Our analysis shed light on the content and targets of these images, finding that Russian trolls had multiple targets: mainly the USA, Ukraine, and Russia.
Furthermore, we found an overall increase in image use after 2016 with a peak in activity during divisive real-world events like the Charlottesville rally.
Finally, by leveraging Hawkes Processes, we quantified the influence that Russian state-sponsored accounts had with respect to the dissemination of images on the Web, finding that these accounts were particularly influential in spreading politics-related images.
Also, by comparing our results to previous analysis made on news URLs, we find that these actors were more influential and efficient in spreading news via URLs when compared to images.

Our findings demonstrate that state-sponsored accounts pursued a political agenda, aimed at influencing users on Web communities w.r.t.~specific world events and individuals (e.g., politicians).
Some of our findings confirm previous analysis performed on the text of the tweets posted by these accounts~\cite{zannettou2018let}, highlighting how state-sponsored actors post images that are conceptually similar to their text, possibly in an attempt to make their content look more credible.
Our influence estimation also demonstrated that Russian state-sponsored accounts were particularly influential and efficient in spreading political images to a handful of Web communities.
Also considering the relatively small number of Russian state-sponsored accounts that were actually identified by Twitter, our analysis suggests that these actors need to be taken very seriously in order to tackle online manipulation and spread of disinformation.

Naturally, our study is not without limitations.
First, our pipeline relies on a closed-system (i.e., Cloud Vision API) with a relatively unknown methodology for extracting entities.
However, our small-scale manual evaluation showed that the API provides an acceptable performance for our needs.
Second, we study the images in isolation, without considering other features of the tweets like shared text, thus we may lose important knowledge that exists in the text like sentiment, entities that are referenced, toxicity, etc.
Finally, our study relies on a dataset that is independently identified and released by Twitter, and the methodology for identifying these accounts is unknown and it is unclear on whether there are false positives within the dataset.

\descr{Implications of our work.}
Overall, our study has several implications related to the exploitation of social media by Russian state-sponsored actors, who share weaponized information on divisive matters with the ultimate goal of sowing discord and influencing online discussions.
As such, their activities should be considered as having broader impact than ``just'' political campaigns, rather, as direct attacks against individuals and communities,
since they can lead to erratic real-world behavior outside the scope of any particular election---e.g., disease epidemics as parents are not vaccinating kids because of disinformation~\cite{broniatowski2018weaponized,mejias2017disinformation}.
We also argue that the public should be adequately informed about the existence and the strategies of these actors, particularly their use of weaponized information beyond just ``fake news,'' as a necessary step toward educating users in how to process and digest information on the Web.

Our analysis also complements, to some extent, the Mueller Report~\cite{muellerreport}.
Although it represents the first comprehensive investigation of large-scale state-sponsored ``information warfare,''
much of the Report currently remains redacted.
Even if it is eventually released in its entirety, it is unlikely to contain a quantitative understanding of how these state-sponsored actors behaved and what kind of influence they had.
Furthermore, state-sponsored attacks are reportedly \emph{still on going}~\cite{trolls_ongoing}.
While still awaiting scientific study, new campaigns, including for instance the Qanon conspiracy theory, have been launched by Russian trolls, and at least partially supported by the use of images initially appearing on imageboards like 4chan and 8chan~\cite{qanon_news}.

Overall, our work can be beneficial to policy makers, law enforcement, and military personnel, as well as political and social scientists, historians, and psychologists who will be studying the events surrounding the 2016 US Presidential Elections for years to come.
Our scientific study of how state sponsored actors used images in their attacks can serve to inform this type of interdisciplinary work by providing, at minimum, a data-backed dissection of the most notable and effective information warfare campaign to date.

Finally, the research community can re-use the tools and techniques presented in this paper to study image sharing by various teams or communities on the Web, e.g., state-sponsored accounts from other countries, bots, or any coordinated campaign.
In fact, Twitter recently released new datasets for state-sponsored trolls that originate from Venezuela and Bangladesh~\cite{new_twitter_dataset}; our techniques can be immediately be applied on this data.

\descr{Future Work.} As part of future work, we plan to study the use of news articles and social network posts from state-sponsored accounts with a particular focus on detecting possibly doctored images.
Finally, we aim to build on top of our work to detect domains that are controlled by state-sponsored actors and aim to push specific (disinformation) narratives on the Web.

\descr{Acknowledgments.} This project has received funding from the European Union's Horizon 2020 Research and Innovation program under the Marie Sk\l{}odowska-Curie ENCASE project (GA No. 691025).
This work reflects only the authors' views and the Commission are not responsible for any use that may be made of the information it contains.

\small
\bibliographystyle{abbrv}

\begin{thebibliography}{10}

\bibitem{badawy2018analyzing}
A.~Badawy, E.~Ferrara, and K.~Lerman.
\newblock {Analyzing the Digital Traces of Political Manipulation: The 2016
  Russian Interference Twitter Campaign}.
\newblock In {\em {ASONAM}}, 2018.

\bibitem{badawy2018falls}
A.~Badawy, K.~Lerman, and E.~Ferrara.
\newblock {Who Falls for Online Political Manipulation?}
\newblock In {\em {WWW Companion}}, 2019.

\bibitem{trolls_ongoing}
J.~Barnes and A.~Goldman.
\newblock {F.B.I. Warns of Russian Interference in 2020 Race and Boosts
  Counterintelligence Operations}.
\newblock
  \url{https://www.nytimes.com/2019/04/26/us/politics/fbi-russian-election-interference.html},
  2019.

\bibitem{berger2012makes}
J.~Berger and K.~L. Milkman.
\newblock {What makes online content viral?}
\newblock {\em Journal of marketing research}, 2012.

\bibitem{blondel2008fast}
V.~D. Blondel, J.-L. Guillaume, R.~Lambiotte, and E.~Lefebvre.
\newblock {Fast Unfolding of Communities in Large Networks}.
\newblock {\em Journal of Statistical Mechanics: Theory and Experiment},
  2008(10), 2008.

\bibitem{boyd2018characterizing}
R.~L. Boyd, A.~Spangher, A.~Fourney, B.~Nushi, G.~Ranade, J.~Pennebaker, and
  E.~Horvitz.
\newblock {Characterizing the Internet Research Agency's Social Media
  Operations During the 2016 US Presidential Election using Linguistic
  Analyses}.
\newblock {\em PsyArXiv}, 2018.

\bibitem{broniatowski2018weaponized}
D.~A. Broniatowski, A.~M. Jamison, S.~Qi, L.~AlKulaib, T.~Chen, A.~Benton,
  S.~C. Quinn, and M.~Dredze.
\newblock {Weaponized health communication: Twitter bots and Russian trolls
  amplify the vaccine debate}.
\newblock {\em American Journal of Public Health}, 108(10), 2018.

\bibitem{charlotesville_death}
C.~Caron.
\newblock {Heather Heyer, Charlottesville Victim, Is Recalled as ``a Strong
  Woman''}.
\newblock \url{https://nyti.ms/2vuxFZx}, 2017.

\bibitem{qanon_news}
B.~Collins and Murphy.
\newblock {Russian troll accounts purged by Twitter pushed Qanon and other
  conspiracy theories }.
\newblock
  \url{https://www.nbcnews.com/tech/social-media/russian-troll-accounts-purged-twitter-pushed-qanon-other-conspiracy-theories-n966091},
  2019.

\bibitem{denning1999information}
D.~E.~R. Denning.
\newblock {\em {Information Warfare and Security}}.
\newblock 1999.

\bibitem{dutt2018senator}
R.~Dutt, A.~Deb, and E.~Ferrara.
\newblock {``Senator, We Sell Ads'': Analysis of the 2016 Russian Facebook Ads
  Campaign}.
\newblock In {\em {International Conference on Intelligent Information
  Technologies}}, 2018.

\bibitem{ester1996density}
M.~Ester, H.-P. Kriegel, J.~Sander, X.~Xu, et~al.
\newblock {A Density-Based Algorithm for Discovering Clusters in Large Spatial
  Databases with Noise}.
\newblock In {\em {KDD}}, 1996.

\bibitem{flores2018mobilizing}
C.~I. Flores-Saviaga, B.~C. Keegan, and S.~Savage.
\newblock Mobilizing the trump train: Understanding collective action in a
  political trolling community.
\newblock In {\em ICWSM}, 2018.

\bibitem{gabielkov2016social}
M.~Gabielkov, A.~Ramachandran, A.~Chaintreau, and A.~Legout.
\newblock {Social clicks: What and who gets read on Twitter?}
\newblock {\em ACM SIGMETRICS Performance Evaluation Review}, 2016.

\bibitem{twitter_russian_iranians_dataset}
V.~Gadde and Y.~Roth.
\newblock {Enabling further research of information operations on Twitter}.
\newblock
  \url{https://blog.twitter.com/official/en_us/topics/company/2018/enabling-further-research-of-information-operations-on-twitter.html},
  2018.

\bibitem{hawkes1971spectra}
A.~G. Hawkes.
\newblock {Spectra of some self-exciting and mutually exciting point
  processes}.
\newblock {\em Biometrika}, 58(1), 1971.

\bibitem{im2019still}
J.~Im, E.~Chandrasekharan, J.~Sargent, P.~Lighthammer, T.~Denby, A.~Bhargava,
  L.~Hemphill, D.~Jurgens, and E.~Gilbert.
\newblock {Still out there: Modeling and Identifying Russian Troll Accounts on
  Twitter}.
\newblock {\em arXiv:1901.11162}, 2019.

\bibitem{jacomy2014forceatlas2}
M.~Jacomy, T.~Venturini, S.~Heymann, and M.~Bastian.
\newblock {ForceAtlas2, a continuous graph layout algorithm for handy network
  visualization designed for the Gephi software}.
\newblock {\em PloS one}, 9(6), 2014.

\bibitem{jenders2013analyzing}
M.~Jenders, G.~Kasneci, and F.~Naumann.
\newblock {Analyzing and predicting viral tweets}.
\newblock In {\em {WWW}}, 2013.

\bibitem{jensen2018russian}
M.~Jensen.
\newblock {Russian Trolls and Fake News: Information or Identity Logics?}
\newblock {\em Journal of International Affairs}, 71(1.5), 2018.

\bibitem{khosla2014makes}
A.~Khosla, A.~Das~Sarma, and R.~Hamid.
\newblock {What makes an image popular?}
\newblock In {\em {Proceedings of the 23rd international conference on World
  wide web}}, 2014.

\bibitem{kim2019tracking}
D.~Kim, T.~Graham, Z.~Wan, and M.-A. Rizoiu.
\newblock {Tracking the Digital Traces of Russian Trolls: Distinguishing the
  Roles and Strategy of Trolls On Twitter}.
\newblock {\em arXiv:1901.05228}, 2019.

\bibitem{koroleva2010stop}
K.~Koroleva, H.~Krasnova, and O.~G{\"u}nther.
\newblock {Stop spamming me!: Exploring information overload on Facebook}.
\newblock In {\em Americas Conference on Information Systems}, 2010.

\bibitem{linderman2014}
S.~W. Linderman and R.~P. Adams.
\newblock {Discovering Latent Network Structure in Point Process Data}.
\newblock In {\em {ICML}}, 2014.

\bibitem{lindermanArxiv}
S.~W. Linderman and R.~P. Adams.
\newblock {Scalable Bayesian Inference for Excitatory Point Process Networks}.
\newblock {\em ArXiv:1507.03228}, 2015.

\bibitem{mejias2017disinformation}
U.~A. Mejias and N.~E. Vokuev.
\newblock {Disinformation and the media: the case of Russia and Ukraine}.
\newblock {\em Media, Culture \& Society}, 39(7), 2017.

\bibitem{monga2006perceptual}
V.~Monga and B.~L. Evans.
\newblock {Perceptual Image Hashing Via Feature Points: Performance Evaluation
  and Tradeoffs}.
\newblock {\em IEEE Transactions on Image Processing}, 2006.

\bibitem{muellerreport}
R.~S. Mueller.
\newblock {Report On The Investigation Into Russian Interference In The 2016
  Presidential Election}.
\newblock US Department of Justice, 2019.

\bibitem{rivers2014ethical}
C.~M. Rivers and B.~L. Lewis.
\newblock {Ethical research standards in a world of big data}.
\newblock {\em F1000Research}, 3, 2014.

\bibitem{new_twitter_dataset}
Y.~Roth.
\newblock {Empowering further research of potential information operations}.
\newblock
  \url{https://blog.twitter.com/en_us/topics/company/2019/further_research_information_operations.html},
  2019.

\bibitem{rowett2018StrategicNeedUnderstand}
G.~Rowett.
\newblock {The Strategic Need to Understand Online Memes and Modern Information
  Warfare Theory}.
\newblock In {\em IEEE Big Data}, 2018.

\bibitem{snopes_kkk}
{Snopes}.
\newblock {Senator Robert Byrd in Ku Klux Klan Garb}.
\newblock \url{https://www.snopes.com/fact-check/robert-byrd-kkk-photo/}, 2016.

\bibitem{spangher2018analysis}
A.~Spangher, G.~Ranade, B.~Nushi, A.~Fourney, and E.~Horvitz.
\newblock {Analysis of Strategy and Spread of Russia-sponsored Content in the
  US in 2017}.
\newblock {\em arXiv:1810.10033}, 2018.

\bibitem{charlotesville}
H.~Spencer.
\newblock {A Far-Right Gathering Bursts Into Brawls}.
\newblock \url{https://nyti.ms/2uTmIgV}, 2017.

\bibitem{stewart2018examining}
L.~G. Stewart, A.~Arif, and K.~Starbird.
\newblock {Examining Trolls and Polarization with a Retweet Network}.
\newblock In {\em {WSDM}}, 2018.

\bibitem{wrzus2013social}
C.~Wrzus, M.~H{\"a}nel, J.~Wagner, and F.~J. Neyer.
\newblock {Social network changes and life events across the life span: a
  meta-analysis}.
\newblock {\em Psychological bulletin}, 2013.

\bibitem{zannettou2018origins}
S.~Zannettou, T.~Caulfield, J.~Blackburn, E.~{De Cristofaro}, M.~Sirivianos,
  G.~Stringhini, and G.~Suarez-Tangil.
\newblock {On the Origins of Memes by Means of Fringe Web Communities}.
\newblock In {\em {ACM IMC}}, 2018.

\bibitem{zannettou2017web}
S.~Zannettou, T.~Caulfield, E.~De~Cristofaro, N.~Kourtellis, I.~Leontiadis,
  M.~Sirivianos, G.~Stringhini, and J.~Blackburn.
\newblock {The Web Centipede: Understanding How Web Communities Influence Each
  Other Through the Lens of Mainstream and Alternative News Sources}.
\newblock In {\em {ACM IMC}}, 2017.

\bibitem{zannettou2018disinformation}
S.~Zannettou, T.~Caulfield, E.~De~Cristofaro, M.~Sirivianos, G.~Stringhini, and
  J.~Blackburn.
\newblock {Disinformation Warfare: Understanding State-Sponsored Trolls on
  Twitter and Their Influence on the Web}.
\newblock In {\em {WWW Companion}}, 2019.

\bibitem{zannettou2018let}
S.~Zannettou, T.~Caulfield, W.~Setzer, M.~Sirivianos, G.~Stringhini, and
  J.~Blackburn.
\newblock {Who Let The Trolls Out? Towards Understanding State-Sponsored
  Trolls}.
\newblock In {\em {WebSci}}, 2019.

\bibitem{Finkelstein2018}
S.~Zannettou, J.~Finkelstein, B.~Bradlyn, and J.~Blackburn.
\newblock {A Quantitative Approach to Understanding Online Antisemitism}.
\newblock In {\em ICWSM}, 2020.

\end{thebibliography}

\end{document}